\documentclass[journal=jpclcd,manuscript=letter]{achemso}
\usepackage[T1]{fontenc}
\usepackage{chemformula}

\usepackage{graphicx}
\usepackage{amssymb}
\usepackage{braket}
\usepackage{mathtools}

\author{Kevin Lively} 
\affiliation{Max Planck Institute for the Structure and Dynamics of Matter and Center for Free-Electron Laser  Science, Luruper Chaussee 149, 22761 Hamburg, Germany}

\author{Guillermo Albareda} 
\affiliation{Max Planck Institute for the Structure and Dynamics of Matter and Center for Free-Electron Laser  Science, Luruper Chaussee 149, 22761 Hamburg, Germany}
\alsoaffiliation{Institute of Theoretical and Computational Chemistry, University of Barcelona, Mart\'i i Franqu\`es 1-11, 08028 Barcelona, Spain}
\alsoaffiliation{Nano-Bio Spectroscopy Group and ETSF, Universidad del País Vasco, 20018 San Sebastían, Spain}

\author{Shunsuke A. Sato}
\affiliation{Max Planck Institute for the Structure and Dynamics of Matter and Center for Free-Electron Laser  Science, Luruper Chaussee 149, 22761 Hamburg, Germany}
\alsoaffiliation{Center for Computational Sciences, University of Tsukuba, Tsukuba 305-8577, Japan}

\author{Aaron Kelly} \email{aaron.kelly@mpsd.mpg.de}
\affiliation{Max Planck Institute for the Structure and Dynamics of Matter and Center for Free-Electron Laser  Science, Luruper Chaussee 149, 22761 Hamburg, Germany}
\alsoaffiliation{Department of Chemistry, Dalhousie University, Halifax B3H 4R2, Canada}

\author{Angel Rubio} \email{angel.rubio@mpsd.mpg.de}
\affiliation{Max Planck Institute for the Structure and Dynamics of Matter and Center for Free-Electron Laser  Science, Luruper Chaussee 149, 22761 Hamburg, Germany}
\alsoaffiliation{Nano-Bio Spectroscopy Group and ETSF, Universidad del País Vasco, 20018 San Sebastían, Spain}
\alsoaffiliation{Center for Computational Quantum Physics (CCQ), Flatiron Institute, 162 Fifth avenue, New York NY 10010, USA}

\title{Simulating Vibronic Spectra without Born-Oppenheimer Surfaces}

\SectionsOff

\begin{document}

\begin{tocentry}
\includegraphics{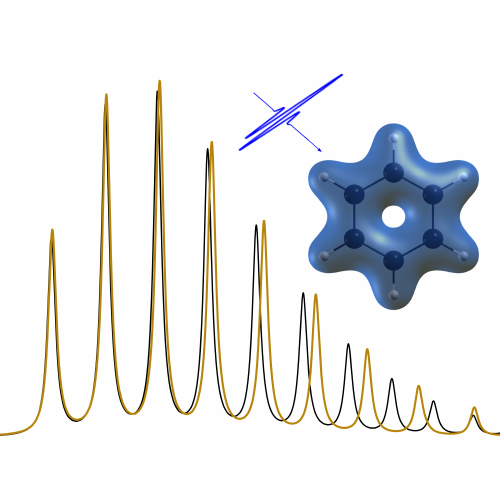}
\end{tocentry}

\begin{abstract}
We show how vibronic spectra in molecular systems can be simulated in an efficient and accurate way using first principles approaches without relying on the explicit use of multiple Born-Oppenheimer potential energy surfaces. We demonstrate and analyse the performance of mean field and beyond mean field dynamics techniques for the \ch{H_2} molecule in one-dimension, in the later case capturing the vibronic structure quite accurately, including quantum Franck-Condon effects. In a practical application of this methodology we simulate the absorption spectrum of benzene in full dimensionality using time-dependent density functional theory at the multi-trajectory mean-field level, finding good qualitative agreement with experiment. These results show promise for future applications of this methodology in capturing phenomena associated with vibronic coupling in more complex molecular, and potentially condensed phase systems.

\end{abstract}

\section{Introduction}

Simulating vibronic effects from first principles calculations is one of the central goals in theoretical spectroscopy that has implications in chemistry, physics and materials science. The involvement of nuclear vibrational quantum states during electronic transitions plays a decisive role in determining the spectral features associated with these processes. This has been well established by the utility of the Franck-Condon principle, for example, which represents an early paradigm for the role of nuclear quantum effects in electronically nonadiabatic processes. Describing this interplay between the electronic and vibrational degrees of freedom requires a quantum mechanical description that is both accurate and scalable to relatively large system sizes. 
One popular method to calculate vibronic spectra is to take a sum-over-states approach, where matrix elements of the transition operators between the states involved in the various processes that generate the desired spectral signal are constructed. In this approach the states of interest are represented using the Born-Oppenheimer (BO) basis; one must already have some \textit{a priori} knowledge of the BO states that are involved, along with the associated potential energy surfaces and nonadiabatic couplings. 

An alternative strategy to the sum over states in the BO basis is to take a coordinate space perspective, and construct the response function for the system of interest from direct time-propagation of the system in that picture\cite{May2011,Ullrich2011}. This invariably requires some level of approximation in the representation dynamics of the electronic and nuclear degrees of freedom, with different consequences for their coupling depending on the method chosen.
The mixed quantum-classical Ehrenfest approach is a practical approximation to the fully quantum mechanical dynamics of the system, and despite it's approximate \textit{dynamics}, a formally exact representation of the quantum \textit{equilibrium} structure of the correlated electronic and vibrational degrees of freedom can be included in a multi-trajectory Ehrenfest (MTEF) simulation through the use of the Wigner representation\cite{Wigner1932,Case2008,Grunwald2009}. In this case, the Wigner transform maps the vibrational quantum states onto phase space distributions of continuous position and momentum coordinates which can be sampled by an appropriate Monte Carlo procedure to capture the quantum equilibrium structure of the problem.
The limitations of the Ehrenfest approach, and other independent trajectory semiclassical methods, are well known\cite{Jasper2004,Karsten2018,Tully1998,Kapral2006,Lee2016} and while there have been many attempts to ameliorate these shortcomings, with some exceptions\cite{Agostini2016,Talotta2020}, most rely on the BO framework in their implementation\cite{Tully1990,Donoso1998,Shalashilin2011,Mignolet2018,Nijjar2019}. In this work we take a different approach to go beyond mean field theory based on the recently introduced interacting conditional wave function (ICWF) formalism, which is able to capture correlated electronic and nuclear dynamics\cite{Albareda2014, Albareda2015,Albareda2016, Albareda2019}. We apply MTEF and ICWF dynamics to an exactly solvable one dimensional \ch{H_2} model, and show that these methods are able to recover electron-nuclear correlations in linear vibronic spectra \textit{without the need to calculate multiple BO surfaces}.  In addition, we show that the MTEF method can be easily extended to \textit{ab initio} non-adiabatic molecular dynamics simulations by calculating the vibronic spectra for benzene, where we find good agreement with experimental results.

\section{Vibronic spectra from linear response}

\subsection{Time Correlation Functions in the Born-Oppenheimer picture}

The linear spectrum of a system is given by the Fourier transform of time correlation function (TCF) $C_{AB}(t) = \braket{[\hat{A}(t),\hat{B}]}$ of the transition dipole operator, $\hat{\mu}$,  $C_{\mu\mu}(t)=\braket{\hat{\mu}(t)\hat{\mu}(0)}$\cite{May2011,Tokmakoff2014}, (unless otherwise stated all expressions are in atomic units): 
\begin{equation}\label{eq:TCF spectra}
    \begin{split}
    I(\omega) &= \frac{4\pi\omega}{3c}\int_{-\infty}^\infty dt\ e^{i\omega t}\braket{\left[\hat{\mu}(t),\hat{\mu}\right]}\\
    &= \frac{8\pi\omega}{3c}\Re\int_0^{\infty} dt\ e^{i\omega t} Tr\Big( \hat{\mu}(t) \hat{\rho}_{eq} \hat{\mu}(t=0) \Big),
\end{split}
\end{equation} where the trace occurs over nuclear and electronic degrees of freedom, $\hat{\rho}_{eq}$ is the equilibrium density matrix for the coupled system, and we evolve $\hat{\mu}(t)$ in the Hilbert representation. 
Traditionally vibronic spectra are explained by invoking the Frank-Condon approximation in the BO picture, where the electronic system is instantly excited, thus promoting the unperturbed ground state nuclear system to a different electronic surface. If one has access to the electronic states involved in a particular spectral range then the contributions to the spectrum due to each electronic transition can be identified by resolving the transition dipole operator in the basis of the electronic states of interest, and the vibronic side peaks of that transition can be calculated by propagating the initial state's nuclear subsystem under the effect of the non-equilibrium electronic occupation. When it is feasible to resolve the nuclear wavefunction dynamics, this can be one of the most accurate methods of calculating molecular vibronic spectra\cite{Raab1999,Vendrell2011}.

\subsection{Kicking the system using a weak perturbing field}
Although resolving eq. (\ref{eq:TCF spectra}) in the BO framework is a powerful analysis tool, it is computationally impractical for systems with many nuclear degrees of freedom, particularly when one desires spectra over multiple surfaces. One can bypass this computational bottleneck by representing the system in a real space basis and using the ``$\delta$-kick" method\cite{Yabana1996}, which captures electronic transitions to all dipole-transition allowed states (resolved on the grid) within a single calculation by utilising the dipole response to a perturbative, but impulsive external field $H_{field} = E(t)\hat{\mu}$, i.e. with $E(t)=\kappa \delta(t)$ and $\kappa << 1$. Using first order perturbation theory, the dipole response $\braket{\Delta\mu(t)} = \braket{\mu(t)}-\braket{\mu(0)}$ can be written in powers of the field\cite{Ullrich2011, May2011},
\begin{equation}
        \braket{\Delta\mu(t)} = i\ Tr\Big( \left[\hat{\mu}^I(t),\hat{\mu}^I(0) \right] \hat{\rho}_{eq}  \Big)\kappa + \mathcal{O}(\kappa^2),
\end{equation}
where $\hat{\mu}^I(t)$ is evolved in the interaction representation.  
Hence, the linear response spectra may also be obtained via the following relation,
\begin{equation}
    C_{\mu\mu}(t) = \frac{-i}{\kappa} \braket{\Delta \mu (t)},
\end{equation} provided the strength of the perturbing field $\kappa$ is sufficiently small. This $\delta$-kick approach only requires the initial state of the full system as input, followed by time propagation for a sufficient interval so as to obtain the desired energy resolution. 
Importantly, this technique can also serve as a foundation for calculating non-linear optical response spectra \cite{DeGiovannini2013}.  

\section{Real Time Dynamics Methods}
While the methods described above are formally equivalent, difference between the calculated spectra can arise when approximations are made. Here we briefly describe two methods for performing coupled electron nuclear dynamics simulations, the quantum-classical mean-field MTEF method, and the ICWF formalism which was designed to go beyond the mean field limit. 

\subsection{Ehrenfest Mean Field Theory}

A typical approach to Ehrenfest theory is to assume a separable electronic-nuclear wave function ansatz, take the classical limit of the nuclear portion, and initialise the nuclei at the equilibrium position with zero nuclear momentum \cite{McLachlan1964, Vacher2016}. This single trajectory Ehrenfest (STEF) method is often employed when a mixed quantum-classical method is needed to couple electronic and nuclear dynamics\cite{Li2005}, in some cases providing a stark difference in electronic dynamics compared to fixed nuclei \cite{AndreaRozzi2013,Krumland2020}. Although attempts at capturing quantized vibrational effects in STEF with the $\delta$-kick method have been made \cite{Goings2016}, they can contain unphysical spectral features (see SI).

An alternative route to Ehrenfest is also possible in the density matrix picture, and proceeds via the quantum-classical Liouville equation \cite{Kapral1999}. 
The major difference is that this representation results in a \textit{multi-trajectory Ehrenfest} picture of the dynamics, where the initial quantum statistics of the correlated system can, in principle, be captured exactly. Here, we outline the evolution equations, and offer more details in the supporting information. The time evolution of the reduced electronic density is \begin{equation}\label{eq:MTEF-subsystem}
    \frac{d}{dt} \hat{\rho}_e(t) = -i\Big[\hat{H}^{Eff}_{e,W}(\mathbf{X}(t)),\ \hat{\rho}_e(t)\Big],
\end{equation} where the subscript $W$ refers to the partial Wigner transform over the nuclei, $\mathbf{X}=(\mathbf{R},\mathbf{P})$ is a collective variable for the nuclear position $\mathbf{R}$ and momentum $\mathbf{P}$, and the effective electronic mean-field Hamiltonian is $\hat{H}^{Eff}_{e,W}(\mathbf{X}(t)) = \hat{H}_{e} + \hat{H}_{en,W}(\mathbf{X}(t))$, where $\hat{H}_{e}$ refers to the electronic portion of the hamiltonian, and $\hat{H}_{en}$ to the electron nuclear coupling. The nuclear dynamics is represented as an ensemble of $N$ independent Wigner phase-space trajectories, $\rho_{n,W} (\mathbf{X},t) = \frac{1}{N}\sum_i^N \delta(\mathbf{X}_i-\mathbf{X}_i(t))$, that evolve according to Hamilton's equations of motion generated from the effective nuclear mean-field Hamiltonian,
\begin{equation}
\begin{split}\label{eq:MTEF-nuclei}
    \frac{\partial \mathbf{R}_{i}}{\partial t} &=  \frac{\partial H_{n,W}^{Eff}}{\partial \mathbf{P}_{i}},\quad \frac{\partial \mathbf{P}_{i}}{\partial t} = - \frac{\partial H_{n,W}^{Eff}}{\partial \mathbf{R}_{i}}\\
    H^{Eff}_{n,W} &= H_{n,W}(\mathbf{R}_i(t)) + Tr_e\Big(\hat{\rho}_e(t)\hat{H}_{en,W}(\mathbf{X}_i(t))\Big).
\end{split}
\end{equation}

The average value of any observable, $\langle O (t) \rangle $, can then be written as follows,
\begin{equation}\label{eq: MTEF-Obs}
\langle O (t) \rangle  = Tr_e \int d\mathbf{X}  \hat{O}_W(\mathbf{X},t) \hat{\rho}_W(\mathbf{X},0)
\end{equation} which can be evaluated by sampling initial conditions from $\hat{\rho}_W(\mathbf{X},0)$, and evolving the expectation value of the observable with according to the above equations of motion. Using this dynamics method in conjunction with the BO basis representation to evaluate eq.s (\ref{eq:TCF spectra}) and (\ref{eq:MTEF-subsystem} - \ref{eq: MTEF-Obs}) ultimately leads to the following equations of motion, with sums over BO states denoted by $a$, (see SI for details)
\begin{equation}\label{eq: MTEF-BO eoms}
\begin{split}
\partial_t\rho_{e}^{aa'} &=-i \rho_{e}^{aa'}(t) (\epsilon_a(\mathbf{R}_i(t)) - \epsilon_{a'}(\mathbf{R}_i(t)))\\
      &+ \sum_{a''}\frac{\mathbf{P}_i(t)}{M}\left(\rho^{aa''}_e(t)d_{a''a'}^i(t) - d_{aa''}^i(t)\rho_e^{a''a'}(t)\right)\\
    \partial_t \mathbf{R}_i(t) &= \mathbf{P}_i(t)/M\\
    \partial_t\mathbf{P}_i(t) &= \sum_{a}-\partial_\mathbf{R}\epsilon_a(\mathbf{R}_i(t))\rho_e^{aa}(t)\\
    &+ \sum_{aa'}\Re\left[\left(\epsilon_a(\mathbf{R}_i(t)) d_{aa'}^i(t) - \epsilon_{a'}(\mathbf{R}_i(t))d_{a'a}^i(t)\right)\rho_e^{a'a}(t)\right]\\
    \partial_t\mu_W^{aa'}(\mathbf{R}_i(t)) &= i\mu_W^{aa'}(\mathbf{R}_i(t))(\epsilon_a(\mathbf{R}_i(t))-\epsilon_{a'}(\mathbf{R}_i(t))).
\end{split}
\end{equation}
Where $\epsilon_a(\mathbf{R})$ are the BO surfaces and $d_{aa'}$ are the non-adiabatic coupling vectors (NACVs) between states $a$ and $a'$. 

In contrast to the previous expression, utilising MTEF in the real space $\delta$-kick approach requires initialising the electronic wave function as the BO eigenstate for each initially sampled nuclear geometry. The $\delta-$kick is applied and the electronic wave function is propagated using the time dependent Schr\"{o}dinger equation equivalent to eq. (\ref{eq:MTEF-subsystem}) alongside the nuclei according to eq. (\ref{eq:MTEF-nuclei}). Calculating the spectrum via MTEF dynamics in the BO picture is from here on referred to as MTEF-BO, and calculating it via the $\delta-$kick method is referred to as MTEF-kick.

\subsection{The Conditional wave function Approach}
Moving beyond semi-classical dynamics, the formally exact CWF method and it's practical ICWF implementation are recently developed methods which have shown to be able to capture non-equilibrium correlated nuclear-nuclear and electron-nuclear phenomena beyond the mean field limit \cite{Albareda2014,Albareda2015,Albareda2016, Albareda2019}. 
This approach is based on taking single-particle slices (the CWFs) of the time-dependent wave function of full system, and approximating the equations of motion for these CWFs by the Hermitian components of the sliced Hamiltonian, and finally, in the ICWF extension, utilising these electron-nuclear CWFs as a basis of Hartree products in a wave function ansatz.

Here we describe an implementation of this approach utilising the static and time-dependent variational principles for the expansion coefficients in a \textit{static} CWF basis. The basis is chosen via sampling electronic and nuclear positions ($\mathbf{r}^{\alpha}, \mathbf{R}^{\alpha}),\ \alpha\in\{1,\ldots, N_c\}$, where $\mathbf{r}$ and $\mathbf{R}$ are understood to be collective position variables, from initial guesses to the electronic and nuclear densities. These are used to construct the Hermitian limit of the CWF propagators\cite{Albareda2014}
\begin{equation}
\begin{split}
    h^{\alpha}_e(\mathbf{r}_i) &= -\frac{1}{2}\nabla^2_{\mathbf{r}_i} + \sum_{j\neq i}^{N_e}V_{ee}(\mathbf{r}_i, \mathbf{r}_j^{\alpha}) +\sum_{l}^{N_n}V_{en}(\mathbf{r}_i,\mathbf{R}^{\alpha}_l) \\
    h^{\alpha}_n(\mathbf{R}_l) &= -\frac{1}{2M_l}\nabla^2_{\mathbf{R}_l} + \sum_{j}^{N_e}V_{en}(\mathbf{R}_l,\mathbf{r}_j^{\alpha}) +\sum_{m\neq l }^{N_n}V_{nn}(\mathbf{R}_l,\mathbf{R}^{\alpha}_m)
\end{split}
\end{equation}
for a system with $N_e$ electrons and $N_n$ nuclear degrees of freedom. Taking eigenstates of $h^{\alpha}_{e}(\mathbf{r}_i)$ and $h^{\alpha}_n(\mathbf{R}_l)$, denoted $\phi^{\alpha}(\mathbf{r}_i)$ and $\chi^{\alpha}(\mathbf{R}_l)$ respectively, as our CWF basis we write the following wave function ansatz:
\begin{equation}
    \Psi(\mathbf{r},\mathbf{R},t) = \sum_{\alpha}^{N_c}C_{\alpha}(t)\prod_i^{N_e}\phi^{\alpha}(\mathbf{r}_i)\prod_l^{N_n}\chi^{\alpha}(\mathbf{R}_l),
\end{equation}
where we have taken a Hartree product of electronic and nuclear CWFs for each degree of freedom. While the Hartree product over electronic degrees of freedom has been sufficient for accuracy in applications of ICWF so far, this ansatz can in principle be trivially extended to have fermionic anti-symmetry via inclusion of Slater determinants. We then utilise the Dirac-Frenkel variational procedure\cite{Broeckhove1988,Lubich2004, Ohta2004} to develop equations of motion for $\vec{C}(t)$, which leads to the following evolution equation for the expansion coefficients,
\begin{equation}\label{eq: ICWF prop}
    \frac{d}{dt}\vec{C} = -i\mathbf{S}^{-1}\mathbf{H}\vec{C},
\end{equation} where
\begin{eqnarray}\nonumber
    \mathbf{S}_{\alpha\beta} &=& \prod_i^{N_e}\int d\mathbf{r}_i(\phi^{\alpha}(\mathbf{r}_i))^*\phi^{\beta}(\mathbf{r}_i)\prod_{l}^{N_n}\int d\mathbf{R}_l(\chi^{\alpha}(\mathbf{R}_l))^*\chi^{\beta}(\mathbf{R}_l),\\\nonumber
    \mathbf{H}_{\alpha\beta} &=& \prod_i^{N_e}\prod_{l}^{N_n}\int d\mathbf{R}_l d\mathbf{r}_i (\phi^{\alpha}(\mathbf{r}_i)\chi^{\alpha}(\mathbf{R}_l))^*\hat{H}(\mathbf{r},\mathbf{R})\phi^{\beta}(\mathbf{r}_i)\chi^{\beta}(\mathbf{R}_l).
\end{eqnarray}
In practice $\mathbf{S}$ may be nearly singular, but its inverse can be approximated by the Moore-Penrose pseudoinverse \cite{Ben-Israel2003}.
The ground state wave function is obtained from this approach using imaginary time evolution\cite{Kosloff1986,Shi2018}, and the $\delta-$kick spectra (ICWF-kick) is calculated by applying the perturbative field to the CWFs at time zero and recalculating the $\mathbf{S}$ and $\mathbf{H}$ matrices, equivalent to propagating in the interaction representation. This "closed-loop" of initial state preparation and time-propagation ensures that our ICWF approach is a fully self-consistent method that increases in accuracy with increasing $N_c$, and requires no BO state information. 

\section{The 1D-Hydrogen molecule}
To investigate the performance of the MTEF and ICWF approaches to vibronic spectral lineshapes we studied the vibronic transitions in an exactly solvable one dimensional model system for molecular Hydrogen \cite{Kreibich2001, Lein2002, Bandrauk2002}.
The total Hamiltonian can be written in the center of mass frame in atomic units as
\begin{equation}
\begin{split}
    \hat{H}(r_1,r_2,R) &= -\frac{\partial^2_R}{2\mu_n} -\sum_{i=1}^2\frac{\partial^2_{r_i}}{2\mu_e}  + \frac{1}{\sqrt{(r_1-r_2)^2 + 1}} + \frac{1}{R} \\
    &-\sum_{i=1}^2 \left(\frac{1}{\sqrt{(r_i + \frac{1}{2} R)^2 +1}} + \frac{1}{\sqrt{(r_i - \frac{1}{2} R)^2 +1}}\right)  
\end{split}
\end{equation}
where $\mu_n=m_p/2$ and $\mu_e=2m_p/(2m_p+1)$ are the reduced nuclear and electronic masses, $R$ is the internuclear separation, and $r_i$ are the electronic positions. We take the proton mass to be $m_p=1836$. The electronic and nuclear degrees of freedom were each resolved on grids for the numerically exact solution and ICWF-kick approaches, while the MTEF-kick electronic wave functions were time evolved on the $(r_1,r_2)$ grid, and the MTEF-BO information was calculated by solving the electronic subsystem across the nuclear grid; see the computational methods section for more details. A kick strength of $\kappa=10^{-4}$a.u. was sufficient to generate the kick spectra within the linear response regime and, unless otherwise stated, a total propagation time of $10,000\ \text{a.u.} \approx 242 fs$ was used to generate the spectra.

\begin{figure}
    \centering
    \includegraphics{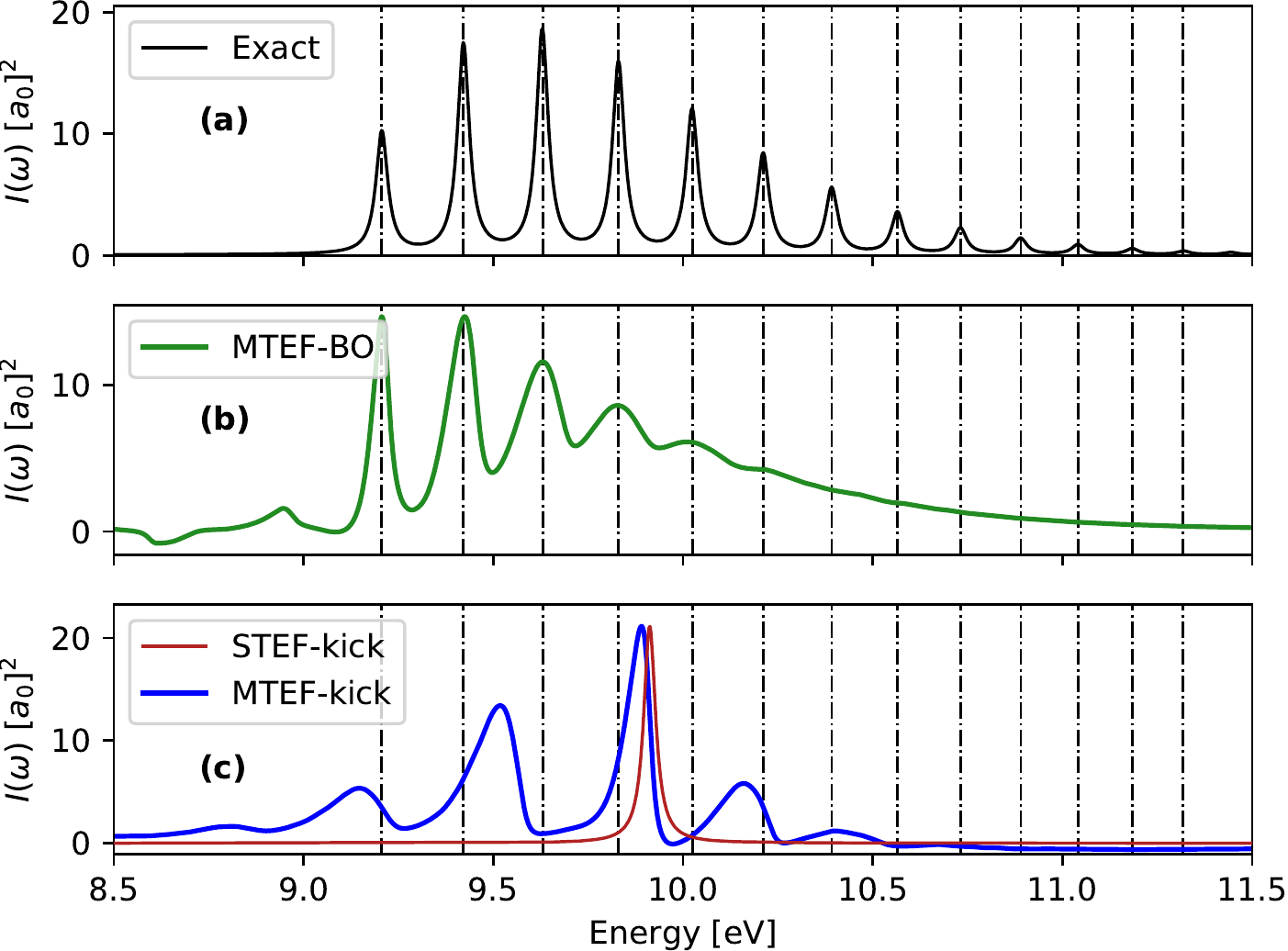}
    \caption{1D \ch{H_2}, $S_2\leftarrow S_0$ spectra calculated via the MTEF-TCF, MTEF-kick and STEF-kick approaches, with the exact peak placements overlaid as dashed vertical lines. Spectral cross sections are reported in square Bohr radii $a_0^2$. For clarity the STEF-kick spectrum has been multiplied by a factor of 0.175 to match the scale of the MTEF-kick results. }
    \label{fig:Absorbtion Spectra}
\end{figure}

In Fig. \ref{fig:Absorbtion Spectra} we show mean field spectra calculated both with (MTEF-BO), and without (MTEF-kick) the use of multiple BO surfaces for the absorption from $S_0$ to $S_2$ in comparison with the numerically exact results.  We see that in the BO picture the MTEF method recovers the vibronic absorbtion peak placement quite accurately for the first five peaks, with a broadening occurring for the higher energy peaks that leads to a loss of structure. This broadening of the spectral signal is due to the well-known fact that the MTEF dynamics does not preserve the correct quantum statistics and thus cannot fully capture the electron-nuclear correlation in the problem (see the SI for a detailed discussion of this issue). The pre-peak features in Fig. \ref{fig:Absorbtion Spectra}b are also unphysical artefacts of MTEF. The MTEF-BO spectra were converged to within graphical accuracy using $N=50,000$ trajectories.

Focusing on the MTEF-kick results in Fig. \ref{fig:Absorbtion Spectra}c, we see that this approach recovers vibronic side peak structures again without any BO surface information, albeit with inaccurate spacing, while STEF-kick captures only the vertical electronic transition from the minimum of the $S_0$ surface. The MTEF-kick spectra converged to within graphical accuracy using $N=30,000$ trajectories.
The average peak spacing in the MTEF-kick spectra is approximately $0.32$eV; this corresponds remarkably well with the natural frequency of the harmonic approximation to the ground state surface expanded around the equilibrium geometry, which is also $0.32$eV in this case. This result is unsurprising as the electronic kick induces a very small population transfer to the upper surface proportional to the square of the kick strength, which results in the mean forces on the nuclei in MTEF-kick essentially corresponding to those of the initial state. 
\begin{figure}
    \centering
    \includegraphics{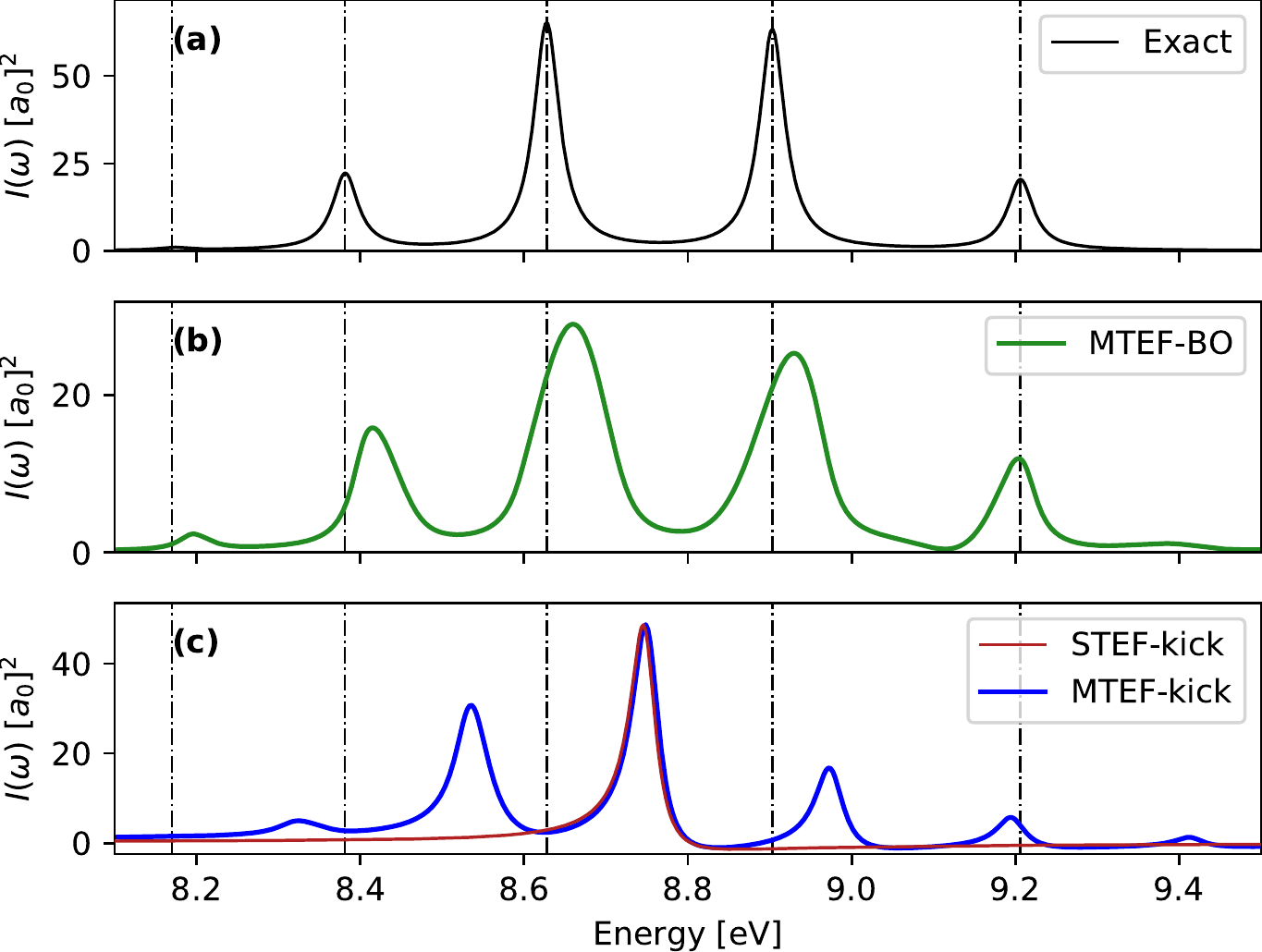}
    \caption{$S_0\leftarrow S_2$ spectra compared between the MTEF-TCF, MTEF-kick and STEF-kick approaches, with exact peak placement overlaid as dashed vertical lines. MTEF nuclear initial conditions are sampled from the lowest lying vibrational state on $S_2$. The sign of all spectra here is inverted for ease of comparison to other figures, and for legibility the STEF-kick spectrum was multiplied by a factor of $0.4$ to match the MTEF-kick spectra maximum. }
    \label{fig:Emission Spectra}
\end{figure}
The influence of initial state on the MTEF-kick spectra is further demonstrated by analysing the emission spectra in Fig. \ref{fig:Emission Spectra}. The initial state here was chosen by hand as the lowest lying nuclear state on the $S_2$ surface. Once again we see that MTEF-BO recovers the peak placement quite well, while the MTEF-kick data has a less accurate vibronic spacing. Fitting the MTEF-kick peaks, we find an excellent correspondence between mean spacing of the five lowest energy MTEF peaks and the excited surface natural frequency of $0.21$eV. 
\begin{figure}
    \centering
    \includegraphics{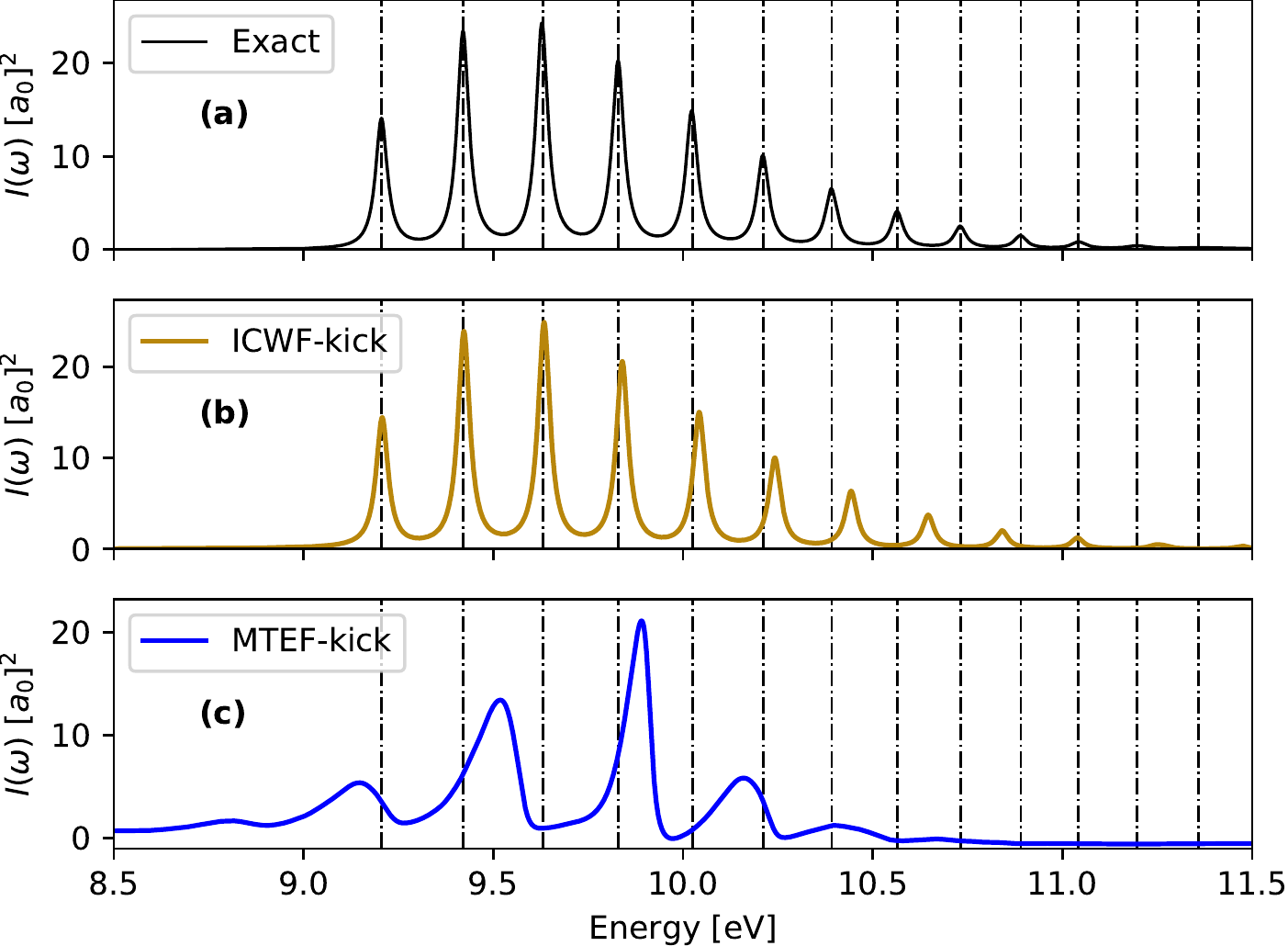}
    \caption{$S_2\leftarrow S_0$ spectra of the ICWF-kick and MTEF-kick methods, with the exact peaks placement overlaid as dashed lines.}
    \label{fig:ICWF absorbtion}
\end{figure}

For ICWF-kick, we found that $N_c=4096$ and mixing the three lowest energy CWF eigenstates was sufficient to obtain quite accurate results. In Fig. \ref{fig:ICWF absorbtion} we demonstrate that the ICWF ansatz used in a variational context achieves a much more accurate vibronic spacing than the MTEF-kick approach, without the failing of peak broadening or unphysical spectral negativity apparent in the MTEF-BO results. The accuracy of these results underscores that the ICWF ansatz is a robust framework to capture the electronic and vibronic quantum dynamics, being accurate for not only the electron-nuclear correlation inherent to vibronic spectra, but also the electronic subsystem itself, which in the MTEF results was solved exactly either on a grid or using explicit BO state information. The deviation from the exact results does grow with increasing energy, although this is ameliorated with increasing $N_c$, and can in principle be eliminated at large enough values of $N_c$ (see SI).

Finally we demonstrate the application of MTEF-kick to real 3D molecular systems using the \textit{ab initio} Octopus\cite{Tancogne-Dejean2020} real-space time dependent density functional theory (TDDFT)\cite{Gross2012} package to calculate the linear vibronic MTEF-kick spectra of Benzene. The initial conditions for the nuclear subsystem were obtained by calculating the normal mode frequencies and dynamical matrix of the molecule, and sampling  Wigner transforms of the ground state wave functions in the harmonic approximation; see SI for more details. The adiabatic-LDA functional was used, along with norm-conserving Troullier-Martins pseudo-potentials, and the trajectories were evolved for $201\frac{\hbar}{eV}\approx 132fs$. A kick strength of $\kappa=5x10^{-3}$a.u. was used to generate the kick spectra within the linear response regime in this case, and the graphical convergence of the MTEF results was found to be achieved with $N=500$ trajectories. 

\section{First principles treatment of Benzene}
\begin{figure}
    \centering
    \includegraphics{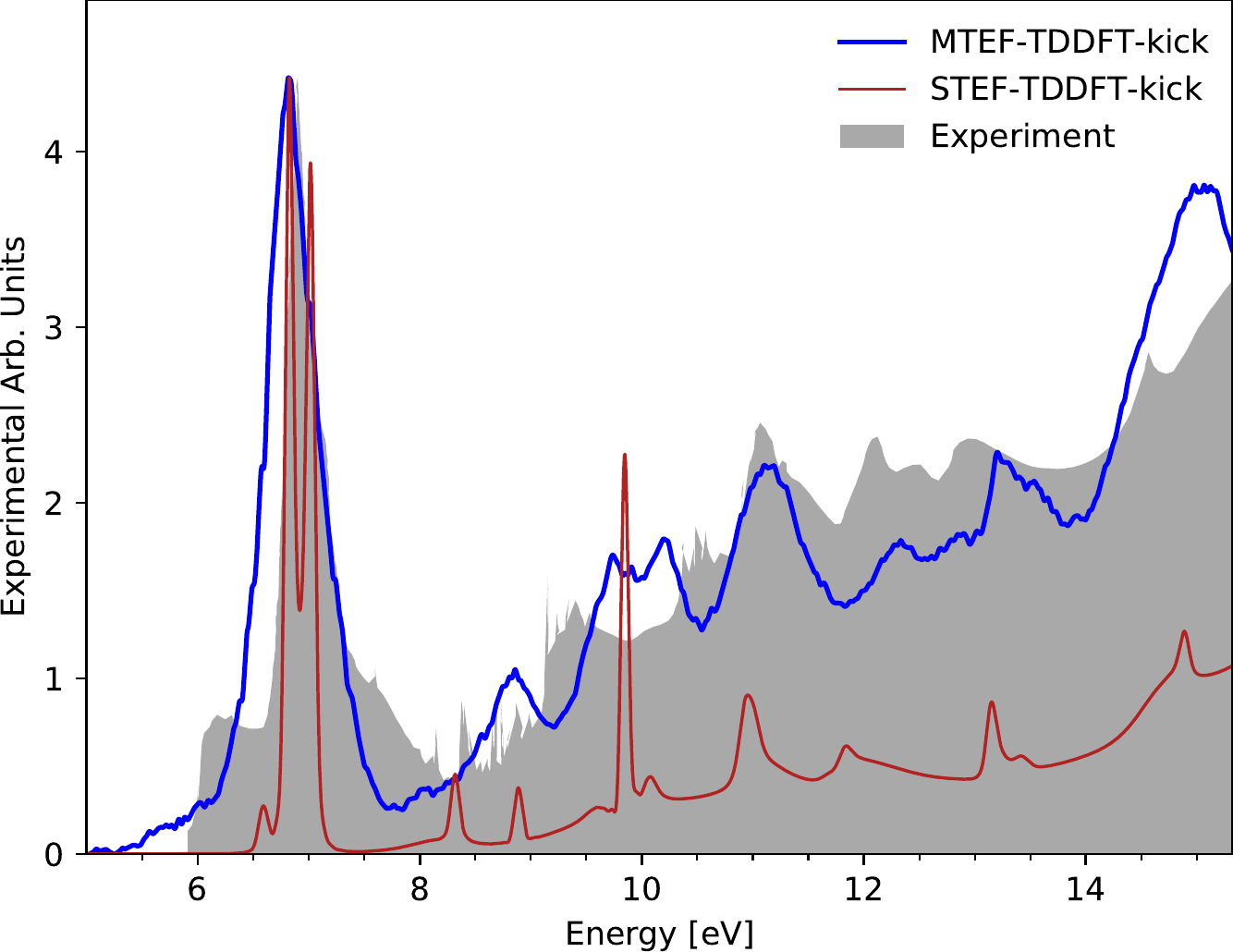}
    \caption{Experimental vibronic spectra for the lowest lying optical transitions of benzene\cite{Koch1972} compared to the MTEF and STEF kick spectra calculated with TDDFT. The spectral weights of the MTEF and STEF data sets have been scaled to the arbitrary units of the experimental data set.}
    \label{fig:Benzene MTEF Kick}
\end{figure}

In Fig. \ref{fig:Benzene MTEF Kick}, we compare the MTEF-TDDFT-kick results to its STEF-TDDFT-kick counterpart, each scaled to match the peak intensity of an experimental data set for the optical absorption of benzene digitized from Ref.~\cite{Koch1972}. We see that there is remarkably good agreement across the wide energy range available from experiment, before molecular dissociation pathways become available around 13.8eV. Again, the full spectrum is resolved without resorting to calculations of individual transitions as would be required in a BO state calculation.
The three STEF peaks in the 7eV region correspond to the energy range of the doubly degenerate, dipole allowed $\prescript{1}{}{E}_{1u}\leftarrow \prescript{1}{}{A}_{1g}$, $\pi^*\leftarrow\pi$ transition\cite{Krumland2020, Koch1972, Gingell1998}, with the energy degeneracy artificially lifted by the discrete grid.
The experimental band preceding the central peak, in the range 6eV to 6.5eV is commonly ascribed to the dipole forbidden, but vibronically allowed $\prescript{1}{}{B}_{1u}\leftarrow \prescript{1}{}{A}_{1g}$ transition\cite{Gingell1998,Koch1972,Borges2003}, and in the MTEF-TDDFT-kick results we see a low energy tail extending through the 5eV-6.25eV range, well away from the STEF results, eventually transitioning into the broad peak centered around 7eV. It's reasonable to expect that the broadening of the MTEF signal relative to the experimental signal is due to the effects discussed above that arise due to the mean field treatment. 

\section{Summary and Outlook}
We have demonstrated that semi-classical MTEF simulations can capture vibronic structure with the correct spectral sign in the region of the transition. Moreover, we have shown how this can be achieved without using multiple BO surfaces via the $\delta$-kick method, and that the vibronic spacing predicted with the MTEF-kick approach matches the profile of the initial state. We have addressed these shortcomings by combining the ICWF formalism with the $\delta$-kick method, which provides more accurate vibronic spectra in a computationally efficient and systematically improvable fashion. Finally, we demonstrated that MTEF-kick is easily applied to \textit{ab initio} molecular systems by simulating the vibronic spectra of benzene and finding good agreement to experimental results. 

These linear response results establish a solid basis for further investigations into non-linear response of field driven molecular systems utilising the practical and efficient MTEF and ICWF techniques along with \textit{ab initio} electronic structure 
methods. Work in preparation by the present authors also explores the utility of ICWF with electron-electron and electron-nuclear correlated systems, and explores the response of these systems under nonperturbative electric fields. Furthermore we expect that MTEF-kick will improve in accuracy for periodic systems, as changes in the electronic configuration are often to likely produce smaller changes in the nuclear forces than in molecular hydrogen. This makes this method interesting to pursue in periodic systems in particular, where there is a dearth of theoretical frameworks for \textit{ab initio}, nonpertubrative electron-nuclear coupling. \cite{Ridolfi2020}
Work in this direction is in progress, as is the implementation of the ICWF method within an \textit{ab initio} framework for molecular and periodic systems.

\section{Computational Methods}
\begin{center}
\textbf{Computational Methods}
\end{center}

In the 1D \ch{H_2} model, the electronic coordinates are each resolved on a $65a_0$ wide interval with spacing $0.6a_0$, while the nuclear grid extends to  $R_{max}=6.3125a_0$ with $0.0625a_0$ spacing. Quadratic complex absorbing potentials were also added to the Hamiltonian to prevent reflection from the simulation box edge (see SI). To generate the exact results we evolved the full wave function under the $\delta-$Kick on the three dimensional electron-nuclear grid, while for MTEF-kick, the electronic subsystem's Schr\"{o}dinger equation, dependent on $\mathbf{R}_i(t)$, was solved exactly on the two dimensional electronic grid for each trajectory. All wave functions were time-propagated using a fourth-order Runge-Kutta integration scheme with a time-step size of $\Delta t=0.05\ \text{a.u.}$. For the MTEF trajectories, the nuclear degree of freedom was propagated via a veloctiy-Verlet type scheme with the same time-step size\cite{Verlet1967}. An exponential damping mask function $\text{exp}(-\gamma t)$ was applied to all time dependent signals in the Fourier transform, with the damping factor was set to damp the signal to 0.1\% it's strength at the final time.

For the 1D \ch{H_2} MTEF-BO results, the potential energy surfaces $\epsilon_a(R)$ and $\mu_W^{aa'}(R)$ were calculated on a nuclear grid with $\Delta R = 0.02a_0$ up to $R_{max}=8a_0$, fit to a cubic spline function, and interpolated every $0.01\Delta R$ \cite{DeBoor2001}. The NACV between $S_0$ and $S_2$ in this model is numerically zero. 
These quantities were resolved for the first allowed dipole transition, between the ground state $S_0$ and the second excited state $S_2$, and the results were found to be well converged within about $5\times 10^4$ trajectories.

For the MTEF-TDDFT-kick simulations we used a real space grid formed from overlapping spheres of radius $8\text{\AA}$ centered on the initial positions of the nuclei, with an isotropic grid spacing of  $0.16\text{\AA}$, which was found to be sufficient to converge the energies of the lowest lying absorption lines.
\begin{acknowledgement}
This work was supported by the European Research Council (ERC-2015-AdG694097), the Cluster of Excellence Advanced Imaging of Matter' (AIM), JSPS KAKENHI Grant Number 20K14382, Grupos Consolidados (IT1249-19) and SFB925. The Flatiron Institute is a division of the Simons Foundation. 
\end{acknowledgement}

\begin{suppinfo}

\section{MTEF Equations of Motion}
\begin{center}
    \textbf{MTEF Equations of Motion}
\end{center}

Starting from a density matrix representation of the full system, $\hat{\rho}$, we Wigner transform over the nuclear subsystem, producing a unique mapping onto a nuclear position $\mathbf{R}$ and momentum $\mathbf{P}$ phase space $\mathbf{X}=(\mathbf{R},\mathbf{P})$, where $\mathbf{R}$ and $\mathbf{P}$ are collective variables $\mathbf{R}=(\mathbf{R}_1,\ldots,\mathbf{R}_{N_n}),\ \mathbf{P}=(\mathbf{P}_1,\ldots,\mathbf{P}_{N_n})$, with $\mathbf{R}_i,\mathbf{P}_i\in\mathbb{R}^d$. The partial wigner transform is defined for any operator as
\begin{equation}
    \hat{\rho}_{W}(\mathbf{R},\mathbf{P}) =  \frac{1}{(2\pi)^{dN_n}}\int d\mathbf{X} e^{i\mathbf{P} \cdot \mathbf{X}} \braket{\mathbf{R}-\frac{\mathbf{X}}{2}|\hat{\rho}|\mathbf{R}+\frac{\mathbf{X}}{2}},
\end{equation}
leaving a Hilbert space operator character over the electronic degrees of freedom, dependent on the continuous nuclear phase space parameters. In general, developing equations of motion for $\hat{\rho}_W(\mathbf{R},\mathbf{P})$, (or any operator), requires taking the partial Wigner transformation of the Liouville von-Neumann equation of motion for $\rho$:
\begin{equation}\label{eq: Wigner Transformed LVN}
    \begin{split}
        \frac{\partial \hat{\rho}_W}{\partial t} &= -i \Big( (\hat{H}\hat{\rho})_W - (\hat{\rho} \hat{H})_W \Big)\\
        (\hat{H}\hat{\rho})_W &= \hat{H}_W\exp\Big( \frac{1}{2i}\Lambda \Big)\hat{\rho}_W\\
        \Lambda &= \overleftarrow{\nabla}_\mathbf{P}\cdot \overrightarrow{\nabla}_\mathbf{R} - \overleftarrow{\nabla}_\mathbf{R}\cdot \overrightarrow{\nabla}_\mathbf{P}\\
        g\exp\Big( \kappa\Lambda \Big)f&= \sum_{s=0}^{\infty} \frac{\kappa^s}{s!}\sum_{t=0}^s (-1)^t {s \choose t}\left[\partial_\mathbf{R}^{s-t}\partial_\mathbf{P}^t\ f\right]\left[\partial_\mathbf{R}^{t}\partial_\mathbf{P}^{s-t}\ g\right].
    \end{split}
\end{equation}
Where the final line defines the ``Moyal product'' also known as the ``star product'' \cite{FAIRLIE1999365}.
By expressing the Poisson braket operator $\Lambda$, in terms of the ratio of masses between the nuclei and the electrons $\Lambda= (m/M)^{\frac{1}{2}}\Lambda'$, and truncating the Moyal product of $e^{(m/M)^{\frac{1}{2}}\Lambda'}$ at first order, one can arrive at the Quantum-Classical Liouville Equation (QCLE):\cite{Kapral1999}
\begin{equation}\label{eq: QCLE}
    i\frac{\partial}{\partial t}\hat{\rho}_W(\mathbf{R},\mathbf{P}) = -i[\hat{H}_W,\hat{\rho}_W] + \frac{1}{2}\Big({\hat{H}_W,\hat{\rho}_W} -{\hat{\rho}_W,\hat{H}_W} \Big),
\end{equation}
where $\{A(\mathbf{R},\mathbf{P}),B(\mathbf{R},\mathbf{P})\}$ refers to the normal Poisson bracket. 

To derive MTEF equations of motion from the QCLE, one takes the mean field approximation by assuming that the full system can be written as a sum of correlated and uncorrelated parts,
\begin{equation}
\hat{\rho}_W(\mathbf{X},t)=\hat{\rho}_e(t) \rho_{n,W}(\mathbf{X},t) + \hat{\rho}_{corr, W}(\mathbf{X},t),
\end{equation} 
and then neglecting the contribution of the correlated part in the \textit{dynamics}. Note that while the ensuing dynamics do not explicitly treat the effect of subsystem correlation, the initial state generally \textit{is} correlated, and therefore is implicitly included in the dynamics.

Under this approximation, the electronic density matrix is
\begin{equation}\label{eq:rdm}
\hat{\rho}_{e} (t) = Tr_n \Big( \hat{\rho}(t) \Big) = \int d\mathbf{X} \hat{\rho}_W (\mathbf{X},t),
\end{equation} and the nuclear (quasi) probability phase space distribution is $\rho_n(\mathbf{X}, t) = Tr_e \left( \hat{\rho}_W (\mathbf{X},t)\right)$.

In the equations of motion resulting from inserting this approximation into the QCLE, the evolution of the reduced Wigner density of the nuclear subsystem can be exactly represented, via the method of characteristics, by a sufficiently large ensemble of multiple independent trajectories, $\rho_{n,W} (\mathbf{X},t) = \frac{1}{N}\sum_i^N \delta(\mathbf{X}_i-\mathbf{X}(t))$. Each trajectory evolves according to Hamilton's equations of motion generated from the mean-field effective Hamiltonian,
\begin{equation}
\begin{split}
    \frac{\partial \mathbf{R}_{i}}{\partial t} &=  \frac{\partial H_{n,W}^{Eff}}{\partial \mathbf{P}_{i}},\quad \frac{\partial \mathbf{P}_{i}}{\partial t} = - \frac{\partial H_{n,W}^{Eff}}{\partial \mathbf{R}_{i}}\\
    H^{Eff}_{n,W} &= H_{n,W}(\mathbf{X}_i(t)) + Tr_e\Big(\hat{H}_{en,W}(\mathbf{X}_i(t))\hat{\rho}^i_e(t)\Big).
\end{split}
\end{equation} Where $H_{n,W}$ and $H_{en,W}$ refer to the partially Wigner transformed nuclear and electron-nuclear coupling operators, respectively. The electronic density associated with each trajectory , $\rho^i_e(t)$, evolves according to the following commutator: 
\begin{equation}\label{eq:EMFT-subsystem}
    \frac{d}{dt} \hat{\rho}^i_e(t) = -i\Big[ \hat{H}_e + \hat{H}_{en,W}(\mathbf{X}_i(t)),\ \hat{\rho}_e^i(t)\Big].
\end{equation}

The exact expression for the average value of any observable, $\langle O (t) \rangle $, can be written as 
\begin{equation}
\begin{split}
\langle O (t) \rangle  &= Tr_e \int d\mathbf{X}  \hat{O}_W(\mathbf{X},t) \hat{\rho}_W(\mathbf{X},0) = Tr_e \int d\mathbf{X}  \hat{O}_W(\mathbf{X}) \hat{\rho}_W(\mathbf{X},t)\\
     &= \sum_i^N Tr_e\Big(\ \hat{O}_W(\mathbf{X}_i(t))\hat{\rho}^i_e(t) \ \Big)
\end{split}
\end{equation} 
The mean field limit of this expression simple corresponds to evaluating the integral by sampling initial conditions for an ensemble of independent trajectories from $\hat{\rho}_W(\mathbf{X},0)$, and then generating the time evolution for each trajectory by approximating $\hat{O}_W(\mathbf{X},t)$ by it's mean-field counterpart. 

Following the sampling of an initial nuclear condition, $\mathbf{X}_i$, from the Wigner distribution associated to the nuclear subsystem wave function, the electronic system is initialised as:
\begin{equation}\label{eq: Electronic Eigenproblem}
    (\hat{H}_e + \hat{H}_{en,W}(\mathbf{R}_i))\phi_a(r) = \epsilon_a(\mathbf{R}_i)\phi_a(r),
\end{equation} i.e. implicitly as the BO electronic state at $\mathbf{R}_i$. Under this scheme, the electronic subsystem's initial conditions are implicitly correlated with the nuclear subsystem's quantum statistics.

In cases where the nuclear initial state is impractical to calculate exactly one may utilise the normal modes of the molecular system, or phonon coordinates of a periodic system, to treat the full nuclear wavefunction as a Hartree product of $N$ uncoupled harmonic oscillators, where $N$ is the number of non-rotational and non-translational nuclear degrees of freedom:
\begin{equation}
\begin{split}
    \chi_n(\mathbf{R}) &\approx \chi_1(Q_1)\otimes\ldots\otimes\chi_N(Q_N)\\
    \chi_i(Q_i) &=\sum_l c^{(i)}_l\chi_i^l(Q_i).
\end{split}
\end{equation}
With $c_l^{(i)}$ referring to the occupation of the $l^{th}$ excited state of normal mode $i$ with wavefunction, $\chi_i^l$, and $Q_i(\mathbf{R})$ the normal mode coordinate. Formally, this is exactly equivalent to taking a second order Taylor expansion approximation of the BO surface about the equilibrium nuclear position $R^0$:
\begin{equation}
\begin{split}
    H_{nuc}(\mathbf{R},\mathbf{P}) = \sum_l \frac{1}{2M_l}\mathbf{P}_l^2 + \sum_{lm}\frac{1}{2}(\mathbf{R}_l-\mathbf{R}_l^0)\frac{\partial^2 V_{BO}}{\partial \mathbf{R}_l\partial \mathbf{R}_m}\bigg|_{\mathbf{R}^0}(\mathbf{R}_m-\mathbf{R}_m^0).
\end{split}
\end{equation}
Defining the dynamical matrix, $\mathcal{H}_{lm} = \frac{1}{\sqrt{M_l}}\frac{\partial^2 V}{\partial \mathbf{R}_l\partial \mathbf{R}_m}\frac{1}{\sqrt{M_m}}$, and it's diagonalizing unitary transform, $D^T\mathcal{H}D = \Omega$, $D^TD=\mathbf{1}$, where $\Omega_{ij} = \omega^2_i\delta_{ij}$, we construct the normal coordinate transform for all non-rotational,  non-translational (imaginary) $\omega^2_{i}$, (here we include $\hbar$ for clarity):
\begin{equation}\label{eq: QtoC}
\begin{split}
    \sqrt{M_l}(\mathbf{R}_l-\mathbf{R}_l^0) &= \sum_{i} D_{li}q_{i}\ , \null\quad \frac{\mathbf{P}_l}{\sqrt{M_l}} = \sum_{i} D_{li}s_{i}\\
    s_{i} &= \sqrt{\hbar\omega_i}S_{i}\ ,\null\quad q_{i} = \sqrt{\frac{\hbar}{\omega_i}}Q_{i},
\end{split}
\end{equation}
such that we obtain the nuclear Hamiltonian in dimensionless normal mode coodinates:
\begin{equation}
    H(Q,S) = \sum_{i} \frac{\hbar\omega_i}{2}(S_{i}^2 + Q_{i}^2).
\end{equation}
Of course, the simple harmonic wave function solutions to the above Hamiltonian have well known analytical expressions and are trivially Wigner transformed, the ground state harmonic oscillator wavefunction's Wigner function for instance is:\cite{Case2008} 
\begin{equation}
        W_0(Q,S) = \frac{1}{\pi}\exp{\left(-S^2 - Q^2\right)}.
\end{equation}
We can therefore sample these transforms for $(Q,S)$ and then use eq. (\ref{eq: QtoC}) to back transform to from normal mode coordinates to cartesian coordinates.

\section{MTEF-BO Equations of Motion in the Born Oppenheimer Basis}
\begin{center}
    \textbf{MTEF-BO Equations of Motion in the Born Oppenheimer Basis}
\end{center}
In deriving the MTEF equations of motion in the BO basis, we start by writing the molecular hamiltonian in terms of position and momentum space operators for the electrons (light particles), $\hat{r},\hat{p}$ and nuclei (heavy particles) $\hat{R},\hat{P}$. These are again understood to be collective variables.
\begin{equation}
\begin{split}
\hat{H}(\hat{r},\hat{p},\hat{R},\hat{P}) &= \frac{1}{2M}\hat{P} ^2+ \hat{h}_e(\hat{r},\hat{p},\hat{R})\\
\hat{h}(\hat{r},\hat{p},\hat{R}) &= \frac{1}{2}\hat{p}^2 + \hat{V}(\hat{r},\hat{R})\\
\hat{V}(\hat{r},\hat{R}) &= \hat{V}_{ee}(\hat{r}) + \hat{V}_{en}(\hat{r},\hat{R}) + \hat{V}_{nn}(\hat{R}).
\end{split}
\end{equation}

We then utilise a position representation in the nuclear dof by expanding in the space of nuclear position states $1_\mathbf{R} = \int d\mathbf{R} \ket{\mathbf{R}}\bra{\mathbf{R}}$, leading to 
\begin{equation}\label{eq:Molecular Hamiltonian}
    \hat{H}(\mathbf{R}) = -\frac{1}{2M}\nabla_{\mathbf{R}}^2 + \hat{h}_e(\hat{r},\hat{p},\mathbf{R})
\end{equation}
For a transition between two electronic states $g$ and $e$,  we can expand in the adiabatic basis $\ket{\phi_a(\mathbf{R})}, (a=g,e)$ which are dependent on the nuclear positions $R$ defined by,
\begin{equation}
\begin{split}
    \hat{h}_e(\mathbf{R})\ket{\phi_a(\mathbf{R})} &= \epsilon_a(\mathbf{R})\ket{\phi_a(\mathbf{R})}.
\end{split}
\end{equation} Taking the partial Wigner transform of eq. (\ref{eq:Molecular Hamiltonian}) leads to 
\begin{equation}
    \hat{H}_W(\mathbf{R},\mathbf{P}) = \frac{1}{2M}\mathbf{P}^2 + \hat{h}_{e,W}(\hat{r},\hat{p},\mathbf{R})
\end{equation}
where $\hat{h}_{e,W}(\mathbf{R})$ is the normal electronic hamiltonian operator, now dependent on $R$ in the Wigner nuclear phase space. 
Starting with the separability approximation for the density operator, and neglecting correlations, we have $\hat{\rho}_W = \hat{\rho}_e\rho_n(\mathbf{R},\mathbf{P})$, with
\begin{equation}
\begin{split}
    \partial_t\hat{\rho}_e &= -i\left[Tr_\mathbf{X}\braket{\hat{h}_{e,W}(\mathbf{R})}, \hat{\rho}_e\right]
\end{split}
\end{equation}
where $Tr_\mathbf{X}\braket{\ldots} = \int \ldots d\mathbf{R}d\mathbf{P} $, and $\mathbf{P}$ scalar terms are cancelled by the commutator. 
We are of course interested in evaluating the dipole-dipole correlation function: 
\begin{equation}
\begin{split}
    C_{\mu\mu}(t) &= \int d\mathbf{R} d\mathbf{P} Tr_{e} \Big\{ \hat{\mu}_W \hat{\sigma}(t) \Big\}\\
    &= \int d\mathbf{R} d\mathbf{P} Tr_{e} \Big\{ \hat{\mu}_W(t) \hat{\sigma}(0) \Big\},
\end{split}
\end{equation}
where $\hat{\sigma} = \left[\hat{\mu}_W,\hat{\rho}_W\right]$, and we resolve the dipole operator as
\begin{equation}
    \begin{split}
        \hat{\mu}_W(\mathbf{R},t=0) &= -\hat{r} + Z_{\mathbf{R}}\mathbf{R}\\
         &= \ket{\phi_a}\braket{\phi_a|-\hat{r}|\phi_{a'}}\bra{\phi_{a'}}  + \delta_{aa'}Z_{\mathbf{R}}\mathbf{R}\\
        &= \begin{pmatrix}
        \mathbf{R} & \mu^{ge}(\mathbf{R})\\
        \mu^{eg}(\mathbf{R}) & \mathbf{R}
        \end{pmatrix}
    \end{split}
\end{equation}
Where $Z_{\mathbf{R}}$ refers to the ionic charge of each nuclei. In practice we can neglect the intra-state $\mathbf{R}$ term as we are focused entirely on the \textit{transition} dipole moment. 

Taking the initial state as the ground state, ($\ket{\Psi} = \ket{\chi_g^0 \phi_g}$)
\begin{equation}
    \hat{\rho}_{W}(\mathbf{R},\mathbf{P}, 0) = \rho_{g}^n(\mathbf{R},\mathbf{P})\begin{pmatrix}
    1 & 0\\
    0 & 0
    \end{pmatrix},
\end{equation}
leads to 
\begin{equation}
\begin{split}
    \hat{\sigma(0)} &= \left[\hat{\mu}_W, \hat{\rho}_W(\mathbf{R},\mathbf{P}, 0)\right]\\
     &= \rho_{g}^n(\mathbf{R},\mathbf{P})\begin{pmatrix}
    0 & -\mu^{ge}(\mathbf{R})\\
    \mu^{eg}(\mathbf{R}) & 0
    \end{pmatrix}.
\end{split}
\end{equation}

And therefore the correlation function becomes
\begin{equation}
\begin{split}
    C_{\mu\mu}(t) &= \int d\mathbf{R} d\mathbf{P}  \left(\mu_W^{ge}(\mathbf{R},t)\sigma^{eg}(0) + \mu_W^{eg}(\mathbf{R},t)\sigma^{ge}(0) \right)\\
    &= \int d\mathbf{R} d\mathbf{P} \big(\mu_W^{ge}(\mathbf{R},t)\mu_W^{eg}(\mathbf{R},0)\\ 
    &- \mu_W^{eg}(\mathbf{R},t)\mu_W^{ge}(\mathbf{R},0)\big)\rho_{g}^n(\mathbf{R},\mathbf{P}).
\end{split}
\end{equation}

We can construct an identical quantity from a different initial condition as a superposition state ($\ket{\Psi} = \frac{1}{\sqrt{2}}\ket{\chi_g}(\ket{\phi_g} + i\ket{\phi_e})$) giving, 
\begin{equation}
    \hat{\tilde{\rho}}_W(\mathbf{R},P,0) = \rho_{g}^n(\mathbf{R},\mathbf{P})\frac{1}{2}\begin{pmatrix}
    1 & -i\\
    i & 1
    \end{pmatrix}
\end{equation}
For this different initial condition we propagate 
\begin{eqnarray}\nonumber
    \tilde{C}_{\mu\mu}(t) &=& \frac{i}{2} \int dR dP \Big( \mu_W^{ge}(\mathbf{R},t)\mu_W^{eg}(\mathbf{R},0) -  \mu_W^{eg}(\mathbf{R},t)\mu_W^{ge}(\mathbf{R},0)\Big) \rho_{g}^n(\mathbf{R},\mathbf{P}) \\
    &=& \frac{i}{2} C_{\mu\mu}(t)
\end{eqnarray}

With this different initial condition, we take the MTEF form of the nuclear density arising from the Monte Carlo integration described above, 
\begin{equation}
    \rho_n(\mathbf{R},\mathbf{P}) = \frac{1}{N}\sum_i \delta(\mathbf{R}-\mathbf{R}_i(t))\delta(\mathbf{P}-\mathbf{P}_i(t)).
\end{equation}
The subsequent equations of motion for the system are for the electronic density, needed for the nuclear trajectories are:
\begin{equation}
\begin{split}
      \partial_t\tilde{\rho}_{e}^{aa'} &=-i \tilde{\rho}_{e}^{aa'}(t) (\epsilon_a(\mathbf{R}_i(t)) - \epsilon_{a'}(\mathbf{R}_i(t)))\\
      &+ \sum_{a''}\frac{\mathbf{P}_i(t)}{M}\left(\tilde{\rho}^{aa''}_e(t)d_{a''a'}^i(t) - d_{aa''}^i(t)\tilde{\rho}_e^{a''a'}(t)\right)\\
    \partial_t \mathbf{R}_i(t) &= \mathbf{P}_i(t)/M\\
    \partial_t\mathbf{P}_i(t) &= \frac{1}{2}\sum_{aa'}\left(F_W^{aa'}(t)\tilde{\rho}^{a'a}_e(t) + \tilde{\rho}^{aa'}_e(t)F_W^{a'a}(t)\right)\\
    &= \sum_{aa'}\Re \left[F_W^{aa'}(t)\tilde{\rho}^{a'a}_e(t)\right]\\
    &= \sum_{a}-\partial_R\epsilon_a(\mathbf{R}_i(t))\tilde{\rho}_e^{aa}(t)\\
    &+ \sum_{aa'}\Re\left[\left(\epsilon_a(\mathbf{R}_i(t)) d_{aa'}^i(t) - \epsilon_{a'}(\mathbf{R}_i(t))d_{a'a}^i(t)\right)\tilde{\rho}_e^{a'a}(t)\right]
\end{split}
\end{equation}
Where in the last two equations we have used the identity $d_{aa'}^i(t) = \braket{\phi_a|\partial_{\mathbf{R}_i}\phi_{a'}}|_{\mathbf{R}_i(t)} = -(d_{a'a}^i(t))^*$, to manipulate $F_W^{aa'}(t) = -\braket{\phi_a(\mathbf{R}) | \partial_R\hat{H}_W|\phi_{a'}(\mathbf{R})}|_{\mathbf{R}_i(t)}$. Note that for transitions like the $S_0/S_2$ transition 1D \ch{H_2} focused on in the body of this paper, the non-adiabatic coupling vector (NACV) $d_{aa'}=0$, means that the mean field force acting on the nuclei is at all times a $\frac{1}{2}$ superposition of the $S_0$ and $S_2$ surfaces.  

These are propagated alongside the dipole matrix element equations of motion, needed for the correlation function:
\begin{equation}
    \partial_t\mu_W^{aa'}(\mathbf{R}_i(t)) = i\mu_W^{aa'}(\mathbf{R}_i(t))(\epsilon_a(\mathbf{R}_i(t))-\epsilon_{a'}(\mathbf{R}_i(t))).
\end{equation}

\section{STEF Spectral Negativity}
\begin{center}
    \textbf{STEF Spectral Negativity}
\end{center}
As mentioned in the main text, previous work by Goings et. al\cite{Goings2016} employed STEF-kick dynamics simulations to calculate spectra in fully ab-initio 3D \ch{H_2} by initialzing the nuclear geometry in non-equilibrium `compressed' geometries. Geometries were selected corresponding to expected vibrational energies from Boltzmann distributions at arbitrary temperatures and the $\delta-$Kick method was used to excite the electronic subsystem. Furthermore, only the \textit{magnitude} of the spectral response was depicted, which does not show the spectral negativity resulting from initialising the mean field simulations in a non-equilibrium state. Here we utilise the canonical initial conditions of the STEF-BO picture for the 1D \ch{H_2} model. The electronic occupation is equal for each of the two surfaces ivolved in the transition, and the nuclear initial condition corresponds to the equillibrium geometry of the initial surface. 
\begin{figure}
    \centering
    \includegraphics{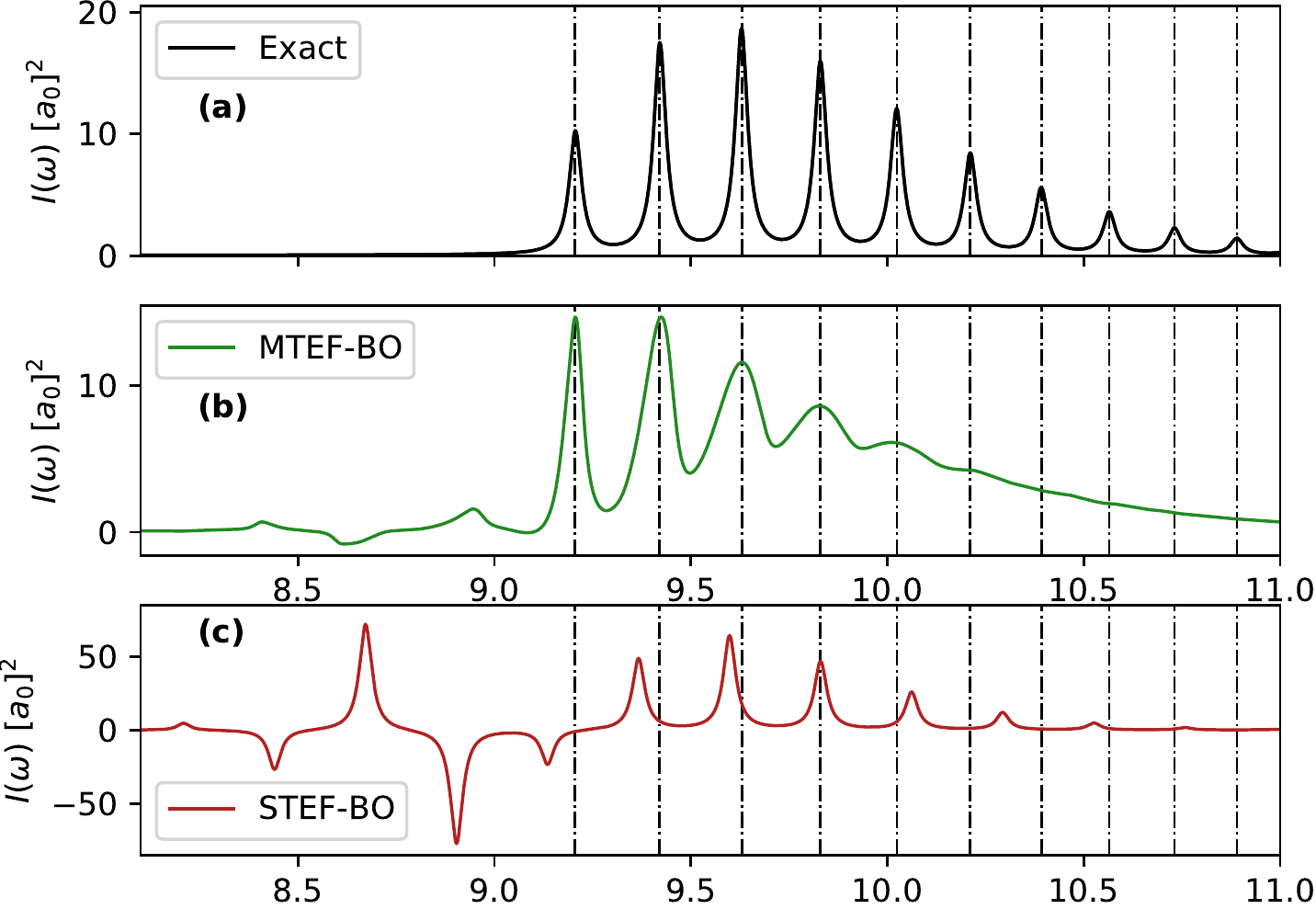}
    \caption{1D \ch{H_2} $S_0\leftarrow S_2$ absorption spectra, comparing exact, MTEF-BO and STEF-BO.}
    \label{fig:SI-STEF-BO Absorbtion}
\end{figure}
In Fig. \ref{fig:SI-STEF-BO Absorbtion}c we see the results of STEF-BO for the $S_2\leftarrow S_0$ region of the spectrum, showing that this only captures positive spectral intensities in the vicinity of the exact results, with accurate peak placement only at the MTEF level. Furthermore the contributions to the unphysical pre-peak features of individual trajectories become apparent in the low energy tail. For completeness we also feature the $S_0\leftarrow S_2$ results in Fig. \ref{fig:SI-STEF-BO Emission}, which demonstrate the same features of correct spectral sign only in the region of the exact results and alternating sign elsewhere. 

\begin{figure}
    \centering
    \includegraphics{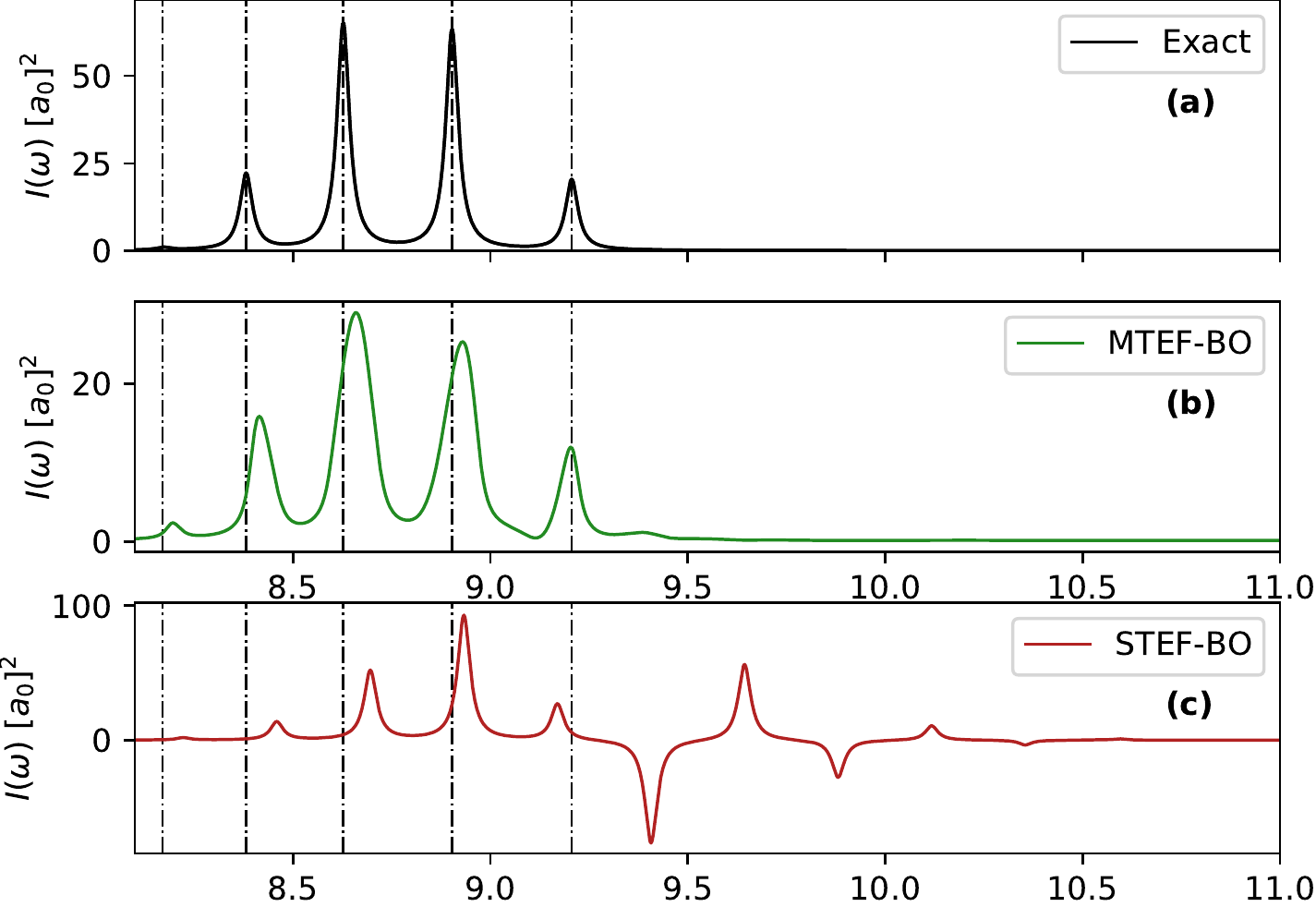}
    \caption{1D \ch{H_2} $S_2\leftarrow S_0$ absorption spectra, comparing exact, MTEF-BO and STEF-BO.}
    \label{fig:SI-STEF-BO Emission}
\end{figure}

\section{Application to Displaced Harmonic Oscillator Model}
\begin{center}
    \textbf{Application to Displaced Harmonic Oscillator Model}
\end{center}
In order to investigate the limitations of MTEF, we can utilise a model which captures the essential physics of the $S_0/S_2$ 1D \ch{H_2} transition which was focused on in the first portion of the main text.
Recall that for this transition, the NACV's between the two electronic adiabatic states are zero, that is $\braket{\phi_a(R) | \partial_{R}\phi_{a'}(R)}=0\ \forall\ a,a'$ in the BO basis, with $a, a'$ restricted to $S_0/S_2$ This means that matrix elements for the partially Wigner transformed molecular hamiltonian can be written as 
\begin{equation}\label{eq: DHO Hamiltonian}
    \begin{split}
        \hat{H}_W(R,P) &= \frac{P^2}{2M} 1 +  \begin{pmatrix}
         \epsilon_g(R) & 0\\
        0 &  \epsilon_e(R)
        \end{pmatrix}.
    \end{split}
\end{equation}
As described in detail in the first section of this SI, MTEF is rooted in a mean field approximation to the QCLE, which is itself the first order expansion of the partially Wigner transformed Liouville von-Neumann equation. Taking eq. (\ref{eq: Wigner Transformed LVN}) to second order provides,
\begin{equation}
    \begin{split}
        \frac{\partial \hat{\rho}_W}{\partial t} &= -i
        \left[ \hat{H}_W, \hat{\rho}_W \right] \\
        &+ \frac{1}{2} \left( \{\hat{H}_W,\hat{\rho}_W\} - \{\hat{\rho}_W,\hat{H}_W\} \right)\\
        &- \frac{i}{8} \left(\left[\partial^2_P \hat{H}_W, \partial^2_{\mathbf{R}}\hat{\rho}_W  \right] + \left[\partial^2_{\mathbf{R}} \hat{H}_W, \partial^2_{\mathbf{P}}\hat{\rho}_W  \right] \right)\\
    \end{split}
\end{equation}
Which in our model Hamiltonian eq. (\ref{eq: DHO Hamiltonian}) becomes, 
\begin{equation}\label{eq: 2nd order LVN}
    \begin{split}
        \frac{\partial \rho^{aa'}_W}{\partial t} &= -i(\epsilon_a(R) - \epsilon_{a'}(R))\rho^{aa'}_W\\
        &+\left[\frac{1}{2}\left(\partial_{R}\epsilon_a(R) + \partial_{R}\epsilon_{a'}(R)\right)\partial_{P} - \frac{P}{M}\partial_{R}\right]\rho^{aa'}_W \\
        &-\frac{i}{8}\left( \partial_{R}^2\epsilon_a(R) - \partial_{R}^2\epsilon_{a'}(R) \right)\partial_{P}^2\rho^{aa'}_W + \mathcal{O}\big((m/M)^{\frac{3}{2}}\big)
    \end{split}
\end{equation}
Such that the error in time propagation resultant from taking only the first order expansion, compared to the second, is proportional to the difference in energy surface curvature. 

If we take the analytically solvable Displaced Harmonic Oscillator (DHO) model\cite{Tokmakoff2014, McKemmish2011} by using surfaces $\epsilon_a(R) = \frac{1}{2}\omega_a^2(R-D_{a})^2 + E_{a}$, we see that for identical surfaces $\omega_e=\omega_g$ that the $2^{\text{nd}}$ order and higher terms in the Wigner transformed Liouville von-Neumann equation are zero, rendering the QCLE exact for this case.

To demonstrate the effect of varying surface curvature, we took parameters similar to harmonic surface fits to the BO surfaces in 1D \ch{H_2}, and for simplicity, took the FC approximation alongside setting $\mu^{aa'}(R) = (1-\delta_{aa'})$a.u..  We solve the exact and MTEF-TCF spectra for the DHO with different values of $\omega_e$ relative to $\omega_g$ by propagating for $T_f=2\cdot 10^4$a.u.. In Fig. (\ref{fig: SI DHO Spectra}) we see iin the left column that for identical upper and lower surfaces, mean field theory is of course exact, and for varying surfaces, MTEF displays a peak broadening and prepeak features. The origin of this broadening is from an effective damping in the time dependent signal, shown in Fig. (\ref{fig: SI DHO Time/Spectra}). 

\begin{figure}
    \centering
    \includegraphics[width=\columnwidth]{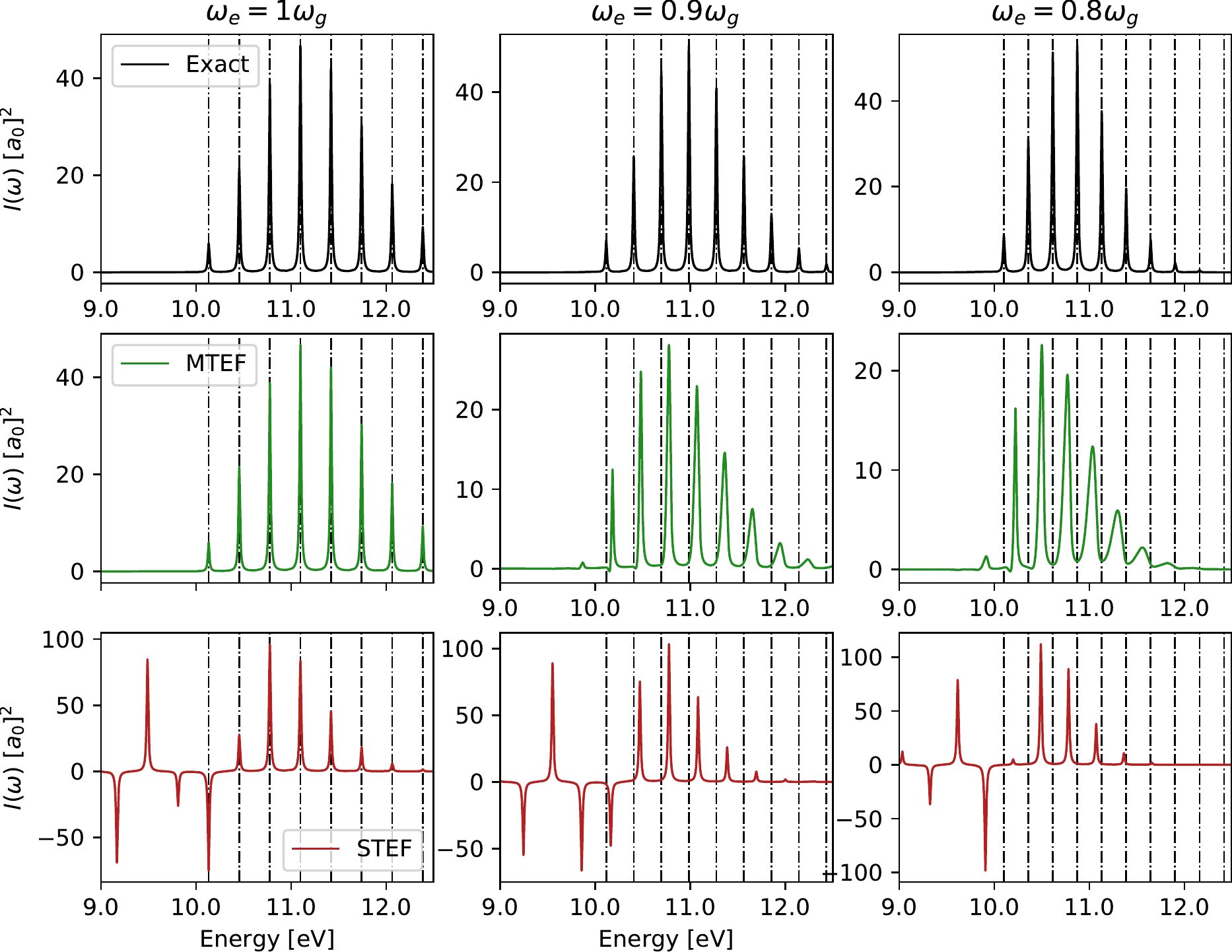}
    \caption{Spectra for the DHO model with several excited and ground state surface frequencies in each column. Each row compares exact, MTEF-BO and STEF-BO results respectively, with the Exact peak placement for each column overlaid across each as vertical dashed lines. }
    \label{fig: SI DHO Spectra}
\end{figure}

\begin{figure}
    \centering
    \includegraphics[width=\columnwidth]{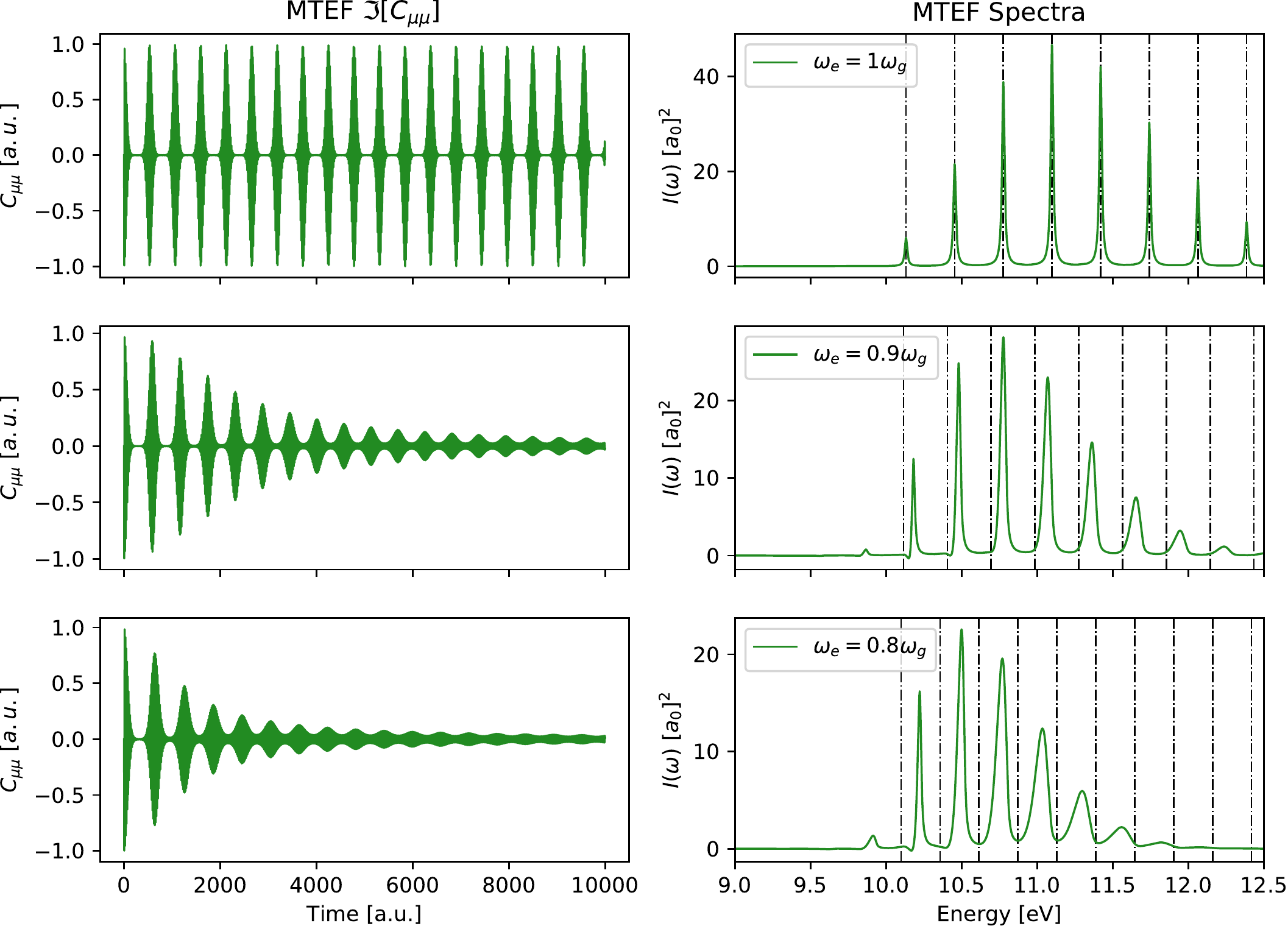}
    \caption{DHO MTEF time dependent dipole-dipole correlation signal in the left column and the resulting spectra in the right column, with the relative surface curvature denoted in the right column legend, and exact spectral peaks overlaid as vertical black dashed lines. For clarity, the time dependent signal is curtailed at $1\cdot 10^4$a.u.. }
    \label{fig: SI DHO Time/Spectra}
\end{figure}

\section{Some More Detail on the ICWF Method}
\begin{center}
    \textbf{Some More Detail on the ICWF Method}
\end{center}
The conditional wave function (CWF) approach can be developed starting from the full molecular wave function for electrons and nuclei, $\Psi(\mathbf{r},\mathbf{R},t)$, which can be formally decomposed in terms of the CWFs of each subsystem:

\begin{eqnarray}\label{CWF1}
	\psi_{e}^\alpha(\mathbf{r},t) &:=&
 \int d\mathbf{R} \delta(\mathbf{R}^\alpha(t) - \mathbf{R}) \Psi(\mathbf{r},\mathbf{R},t),	
\\ \label{CWF2}
	\psi_{n}^\alpha(\mathbf{R},t) &:=& 
 \int d\mathbf{r} \delta(\mathbf{r}^\alpha(t) - \mathbf{r}) \Psi(\mathbf{r},\mathbf{R},t).
\end{eqnarray}

From these definitions one can show that the CWFs, $\psi_{e}^\alpha(t)$ and $\psi_{n}^\alpha(t)$, obey non-Hermitian equations of motion
involving complex potentials which are functionals of the full wave function
and cause the time-evolution of the individual CWFs to be non-unitary\cite{Albareda2014}.
The recently developed Interacting-CWF (ICWF) method\cite{Albareda2019} avoids the direct calculation of these nonlocal complex potentials by positing the following multiconfigurational CWF basis ansatz for the full many-body wave function: 
\begin{equation}\label{ansatz}
	\Psi(r,\mathbf{R},t) = \sum_{\alpha=1}^{N_c} C_\alpha(t) \psi_{e}^\alpha(\mathbf{r},t)
    \psi_{n}^\alpha(\mathbf{R},t).
\end{equation} The basis functions in this sum are chosen to be single particle CWFs that satisfy the mean-field, or Hermitian, limit of the CWF equations in which the complex potentials trivially vanish. The upper limit of the sum, $N_c$, refers to the total number of configurations, which can be stochastically sampled. Including interactions between the trajectories in the ensemble through the coefficients $\mathbf{C}(t) = \left\{ C_1(t),...,C_{N_c}(t) \right\}$ corrects the Hermitian-CWF evolution. The time evolution of these coefficients is obtained by inserting eq. (\ref{ansatz}) directly into the TDSE. 

As described in the text, for the kick spectra adapted ICWF algorithm, the CWFs are instead selected as eigenstates of the Hermitian propagators, and used as a static basis. The imaginary and real time equations of motion for the expansion coefficient $\vec{C}$ are then solved using the respective variational principles\cite{Shi2018,Broeckhove1988,Lubich2004,Ohta2004}, allowing for a completely closed-loop algorithm for wave function preparation and propagation.

To generate the kick spectra, after preparing the ground state $\vec{C}(0)$, the relevant degree of freedom of the kick operator $\text{exp}(-i\kappa\hat{\mu})$ is applied to each CWF, the Hamiltonian and inverse overlap matrices are reconstructed, and $\vec{C}$ is propagated to the desired time. This procedure is equivalent to propagating in the interaction representation, with $\hat{V}_{I}(t) = \kappa \delta(t)\hat{\mu}$. Since these matrices are only constructed at time zero, this algorithm is extremely efficient, requiring only the propagation of a $N_c\times 1$ vector by a $N_c \times N_c$ matrix. For comparison, the 1D H2 MTEF-kick results reported here required the propagation of $34,000$ trajectories each consisting of $108^2\times 1$ electronic wave functions. With a parallelized implementation and hardware allowing approximately $50\text{traj/hr}$, this equates to roughly $680$ compute hours. The ICWF $N_c=4096$ results reported in the main body by contrast require $17$ compute hours on the same hardware.

With increasing non-redundant variational parameters, one is guaranteed to better capture the initial state and minimize the error of time dependent propagation\cite{Lubich2004}. As an example of the convergence properties of ICWF-kick, see Fig. \ref{fig: SI-ICWF Convergence}. These spectra are the result of utilising only lowest energy hermitian propagator eigenstates and propagating for $T_f=1500$ a.u.  with a mask function\cite{Yabana2006} $W(x) = 1 -3x^2 + 2x^3$, for $x=t/T_f$ applied to the time signal in the Fourier Transform. The more accurate $N_c=4096$ results in the main body are initialised using mixes of various excited eigenstates of the propagators.
Theoretical and practical developments are underway to implement this method in arbitrary ab-initio settings. 

\begin{figure}
    \centering
    \includegraphics[width=\columnwidth]{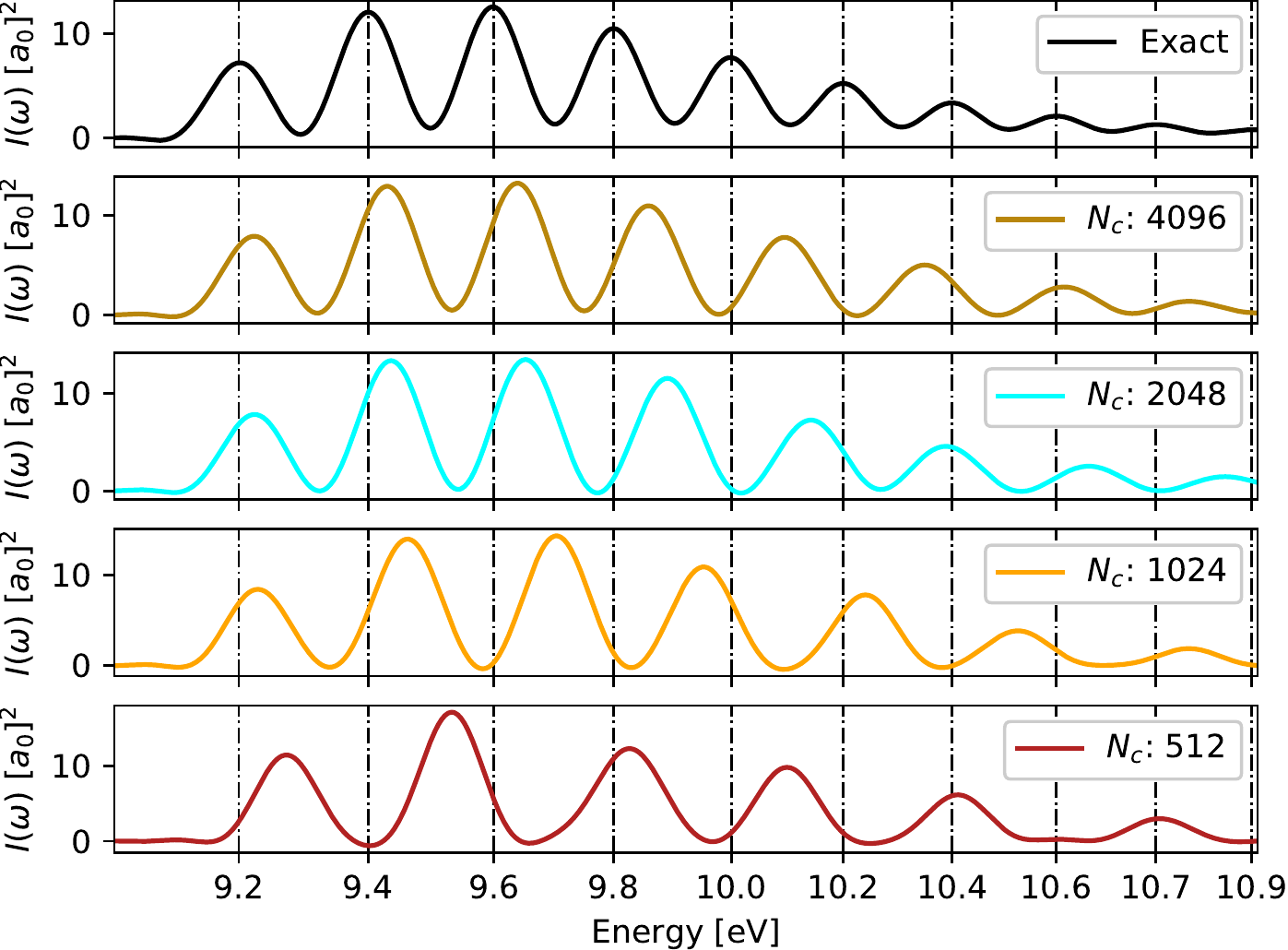}
    \caption{Convergence of the 1D \ch{H_2} spectra for ICWF-kick with different numbers ($N_c$) of lowest energy eigenstate CWF bases.}
    \label{fig: SI-ICWF Convergence}
\end{figure}

\section{Complex Absorbing Potentials}
\begin{center}
    \textbf{Complex Absorbing Potentials}
\end{center}
Quadratic complex absorbing potentials\cite{Muga2004} of the following form were used in all simulations of the one dimensional \ch{H_2} model:
\begin{equation}
\begin{split}
    W_e(r_i) &= -i\eta \left[(r_i-r_l)^2\Theta(r_l-r_i) + (r_i-r_r)^2\Theta(r_i-r_r)\right]\\
    W_n(R) &= -i\eta(R-R_r)^2\Theta(R-R_0),
\end{split}
\end{equation}
where $\Theta$ is the Heaviside function, and $\eta$ was set to 0.1Ha/$a_0$ for both subsystems. 

The electronic CAP cut offs, $r_l$ and $r_r$, were placed $10a_0$ from the walls, while the nuclear CAP start was set at $R_0=5.6875a_0$.
\end{suppinfo}
\bibliography{main}

\providecommand{\latin}[1]{#1}
\makeatletter
\providecommand{\doi}
  {\begingroup\let\do\@makeother\dospecials
  \catcode`\{=1 \catcode`\}=2 \doi@aux}
\providecommand{\doi@aux}[1]{\endgroup\texttt{#1}}
\makeatother
\providecommand*\mcitethebibliography{\thebibliography}
\csname @ifundefined\endcsname{endmcitethebibliography}
  {\let\endmcitethebibliography\endthebibliography}{}
\begin{mcitethebibliography}{55}
\providecommand*\natexlab[1]{#1}
\providecommand*\mciteSetBstSublistMode[1]{}
\providecommand*\mciteSetBstMaxWidthForm[2]{}
\providecommand*\mciteBstWouldAddEndPuncttrue
  {\def\EndOfBibitem{\unskip.}}
\providecommand*\mciteBstWouldAddEndPunctfalse
  {\let\EndOfBibitem\relax}
\providecommand*\mciteSetBstMidEndSepPunct[3]{}
\providecommand*\mciteSetBstSublistLabelBeginEnd[3]{}
\providecommand*\EndOfBibitem{}
\mciteSetBstSublistMode{f}
\mciteSetBstMaxWidthForm{subitem}{(\alph{mcitesubitemcount})}
\mciteSetBstSublistLabelBeginEnd
  {\mcitemaxwidthsubitemform\space}
  {\relax}
  {\relax}

\bibitem[May and K{\"{u}}hn(2011)May, and K{\"{u}}hn]{May2011}
May,~V.; K{\"{u}}hn,~O. \emph{Charge and Energy Transfer Dynamics in Molecular
  Systems: Third Edition}; 2011\relax
\mciteBstWouldAddEndPuncttrue
\mciteSetBstMidEndSepPunct{\mcitedefaultmidpunct}
{\mcitedefaultendpunct}{\mcitedefaultseppunct}\relax
\EndOfBibitem
\bibitem[Ullrich(2011)]{Ullrich2011}
Ullrich,~C.~A. {Time-Dependent Density-Functional Theory: Concepts and
  Applications}. \emph{Oxford Graduate Texts} \textbf{2011}, \relax
\mciteBstWouldAddEndPunctfalse
\mciteSetBstMidEndSepPunct{\mcitedefaultmidpunct}
{}{\mcitedefaultseppunct}\relax
\EndOfBibitem
\bibitem[Wigner(1932)]{Wigner1932}
Wigner,~E. {On the quantum correction for thermodynamic equilibrium}.
  \emph{Physical Review} \textbf{1932}, \relax
\mciteBstWouldAddEndPunctfalse
\mciteSetBstMidEndSepPunct{\mcitedefaultmidpunct}
{}{\mcitedefaultseppunct}\relax
\EndOfBibitem
\bibitem[Case(2008)]{Case2008}
Case,~W.~B. {Wigner functions and Weyl transforms for pedestrians}.
  \emph{American Journal of Physics} \textbf{2008}, \emph{76}, 937--946\relax
\mciteBstWouldAddEndPuncttrue
\mciteSetBstMidEndSepPunct{\mcitedefaultmidpunct}
{\mcitedefaultendpunct}{\mcitedefaultseppunct}\relax
\EndOfBibitem
\bibitem[Grunwald \latin{et~al.}(2009)Grunwald, Kelly, and
  Kapral]{Grunwald2009}
Grunwald,~R.; Kelly,~A.; Kapral,~R. \emph{{Quantum Dynamics in Almost Classical
  Environments}}; 2009\relax
\mciteBstWouldAddEndPuncttrue
\mciteSetBstMidEndSepPunct{\mcitedefaultmidpunct}
{\mcitedefaultendpunct}{\mcitedefaultseppunct}\relax
\EndOfBibitem
\bibitem[Jasper \latin{et~al.}(2004)Jasper, Zhu, Nangia, and
  Truhlar]{Jasper2004}
Jasper,~A.~W.; Zhu,~C.; Nangia,~S.; Truhlar,~D.~G. {Introductory lecture:
  Nonadiabatic effects in chemical dynamics}. Faraday Discussions. 2004\relax
\mciteBstWouldAddEndPuncttrue
\mciteSetBstMidEndSepPunct{\mcitedefaultmidpunct}
{\mcitedefaultendpunct}{\mcitedefaultseppunct}\relax
\EndOfBibitem
\bibitem[Karsten \latin{et~al.}(2018)Karsten, Ivanov, Bokarev, and
  K{\"{u}}hn]{Karsten2018}
Karsten,~S.; Ivanov,~S.~D.; Bokarev,~S.~I.; K{\"{u}}hn,~O. {Quasi-classical
  approaches to vibronic spectra revisited}. \emph{Journal of Chemical Physics}
  \textbf{2018}, \relax
\mciteBstWouldAddEndPunctfalse
\mciteSetBstMidEndSepPunct{\mcitedefaultmidpunct}
{}{\mcitedefaultseppunct}\relax
\EndOfBibitem
\bibitem[Tully(1998)]{Tully1998}
Tully,~J.~C. {Mixed quantum-classical dynamics}. \emph{Faraday Discussions}
  \textbf{1998}, \relax
\mciteBstWouldAddEndPunctfalse
\mciteSetBstMidEndSepPunct{\mcitedefaultmidpunct}
{}{\mcitedefaultseppunct}\relax
\EndOfBibitem
\bibitem[Kapral(2006)]{Kapral2006}
Kapral,~R. {Progress in the theory of mixed quantum-classical dynamics}. Annual
  Review of Physical Chemistry. 2006\relax
\mciteBstWouldAddEndPuncttrue
\mciteSetBstMidEndSepPunct{\mcitedefaultmidpunct}
{\mcitedefaultendpunct}{\mcitedefaultseppunct}\relax
\EndOfBibitem
\bibitem[Lee \latin{et~al.}(2016)Lee, Huo, and Coker]{Lee2016}
Lee,~M.~K.; Huo,~P.; Coker,~D.~F. {Semiclassical Path Integral Dynamics:
  Photosynthetic Energy Transfer with Realistic Environment Interactions}.
  \emph{Annual Review of Physical Chemistry} \textbf{2016}, \relax
\mciteBstWouldAddEndPunctfalse
\mciteSetBstMidEndSepPunct{\mcitedefaultmidpunct}
{}{\mcitedefaultseppunct}\relax
\EndOfBibitem
\bibitem[Agostini \latin{et~al.}(2016)Agostini, Min, Abedi, and
  Gross]{Agostini2016}
Agostini,~F.; Min,~S.~K.; Abedi,~A.; Gross,~E. K.~U. {Quantum-Classical
  Nonadiabatic Dynamics: Coupled- vs Independent-Trajectory Methods}.
  \emph{Journal of Chemical Theory and Computation} \textbf{2016}, \emph{12},
  2127--2143\relax
\mciteBstWouldAddEndPuncttrue
\mciteSetBstMidEndSepPunct{\mcitedefaultmidpunct}
{\mcitedefaultendpunct}{\mcitedefaultseppunct}\relax
\EndOfBibitem
\bibitem[Talotta \latin{et~al.}(2020)Talotta, Agostini, and
  Ciccotti]{Talotta2020}
Talotta,~F.; Agostini,~F.; Ciccotti,~G. {Quantum Trajectories for the Dynamics
  in the Exact Factorization Framework: A Proof-of-Principle Test}. \emph{The
  Journal of Physical Chemistry A} \textbf{2020}, \emph{124}, 6764--6777\relax
\mciteBstWouldAddEndPuncttrue
\mciteSetBstMidEndSepPunct{\mcitedefaultmidpunct}
{\mcitedefaultendpunct}{\mcitedefaultseppunct}\relax
\EndOfBibitem
\bibitem[Tully(1990)]{Tully1990}
Tully,~J.~C. {Molecular dynamics with electronic transitions}. \emph{The
  Journal of Chemical Physics} \textbf{1990}, \relax
\mciteBstWouldAddEndPunctfalse
\mciteSetBstMidEndSepPunct{\mcitedefaultmidpunct}
{}{\mcitedefaultseppunct}\relax
\EndOfBibitem
\bibitem[Donoso and Martens(1998)Donoso, and Martens]{Donoso1998}
Donoso,~A.; Martens,~C.~C. {Simulation of Coherent Nonadiabatic Dynamics Using
  Classical Trajectories}. \emph{The Journal of Physical Chemistry A}
  \textbf{1998}, \emph{102}, 4291--4300\relax
\mciteBstWouldAddEndPuncttrue
\mciteSetBstMidEndSepPunct{\mcitedefaultmidpunct}
{\mcitedefaultendpunct}{\mcitedefaultseppunct}\relax
\EndOfBibitem
\bibitem[Shalashilin(2011)]{Shalashilin2011}
Shalashilin,~D.~V. {Multiconfigurational Ehrenfest approach to quantum coherent
  dynamics in large molecular systems}. \emph{Faraday Discussions}
  \textbf{2011}, \relax
\mciteBstWouldAddEndPunctfalse
\mciteSetBstMidEndSepPunct{\mcitedefaultmidpunct}
{}{\mcitedefaultseppunct}\relax
\EndOfBibitem
\bibitem[Mignolet and Curchod(2018)Mignolet, and Curchod]{Mignolet2018}
Mignolet,~B.; Curchod,~B.~F. {A walk through the approximations of ab initio
  multiple spawning}. \emph{Journal of Chemical Physics} \textbf{2018}, \relax
\mciteBstWouldAddEndPunctfalse
\mciteSetBstMidEndSepPunct{\mcitedefaultmidpunct}
{}{\mcitedefaultseppunct}\relax
\EndOfBibitem
\bibitem[Nijjar \latin{et~al.}(2019)Nijjar, Jankowska, and Prezhdo]{Nijjar2019}
Nijjar,~P.; Jankowska,~J.; Prezhdo,~O.~V. {Ehrenfest and classical path
  dynamics with decoherence and detailed balance}. \emph{Journal of Chemical
  Physics} \textbf{2019}, \relax
\mciteBstWouldAddEndPunctfalse
\mciteSetBstMidEndSepPunct{\mcitedefaultmidpunct}
{}{\mcitedefaultseppunct}\relax
\EndOfBibitem
\bibitem[Albareda \latin{et~al.}(2014)Albareda, Appel, Franco, Abedi, and
  Rubio]{Albareda2014}
Albareda,~G.; Appel,~H.; Franco,~I.; Abedi,~A.; Rubio,~A. {Correlated
  electron-nuclear dynamics with conditional wave functions}. \emph{Physical
  Review Letters} \textbf{2014}, \relax
\mciteBstWouldAddEndPunctfalse
\mciteSetBstMidEndSepPunct{\mcitedefaultmidpunct}
{}{\mcitedefaultseppunct}\relax
\EndOfBibitem
\bibitem[Albareda \latin{et~al.}(2015)Albareda, Bofill, Tavernelli,
  Huarte-Larranaga, Illas, and Rubio]{Albareda2015}
Albareda,~G.; Bofill,~J.~M.; Tavernelli,~I.; Huarte-Larranaga,~F.; Illas,~F.;
  Rubio,~A. {Conditional born-oppenheimer dynamics: Quantum dynamics
  simulations for the model porphine}. \emph{Journal of Physical Chemistry
  Letters} \textbf{2015}, \relax
\mciteBstWouldAddEndPunctfalse
\mciteSetBstMidEndSepPunct{\mcitedefaultmidpunct}
{}{\mcitedefaultseppunct}\relax
\EndOfBibitem
\bibitem[Albareda \latin{et~al.}(2016)Albareda, Abedi, Tavernelli, and
  Rubio]{Albareda2016}
Albareda,~G.; Abedi,~A.; Tavernelli,~I.; Rubio,~A. {Universal steps in quantum
  dynamics with time-dependent potential-energy surfaces: Beyond the
  Born-Oppenheimer picture}. \emph{Physical Review A} \textbf{2016}, \relax
\mciteBstWouldAddEndPunctfalse
\mciteSetBstMidEndSepPunct{\mcitedefaultmidpunct}
{}{\mcitedefaultseppunct}\relax
\EndOfBibitem
\bibitem[Albareda \latin{et~al.}(2019)Albareda, Kelly, and Rubio]{Albareda2019}
Albareda,~G.; Kelly,~A.; Rubio,~A. {Nonadiabatic quantum dynamics without
  potential energy surfaces}. \emph{Physical Review Materials} \textbf{2019},
  \relax
\mciteBstWouldAddEndPunctfalse
\mciteSetBstMidEndSepPunct{\mcitedefaultmidpunct}
{}{\mcitedefaultseppunct}\relax
\EndOfBibitem
\bibitem[Tokmakoff(2014)]{Tokmakoff2014}
Tokmakoff,~A. {Time-Dependent Quantum Mechanics and Spectroscopy}.
  \emph{Lecture} \textbf{2014}, \relax
\mciteBstWouldAddEndPunctfalse
\mciteSetBstMidEndSepPunct{\mcitedefaultmidpunct}
{}{\mcitedefaultseppunct}\relax
\EndOfBibitem
\bibitem[Raab \latin{et~al.}(1999)Raab, Worth, Meyer, and Cederbaum]{Raab1999}
Raab,~A.; Worth,~G.~A.; Meyer,~H.-D.; Cederbaum,~L.~S. {Molecular dynamics of
  pyrazine after excitation to the S2 electronic state using a realistic
  24-mode model Hamiltonian}. \emph{The Journal of Chemical Physics}
  \textbf{1999}, \relax
\mciteBstWouldAddEndPunctfalse
\mciteSetBstMidEndSepPunct{\mcitedefaultmidpunct}
{}{\mcitedefaultseppunct}\relax
\EndOfBibitem
\bibitem[Vendrell and Meyer(2011)Vendrell, and Meyer]{Vendrell2011}
Vendrell,~O.; Meyer,~H.~D. {Multilayer multiconfiguration time-dependent
  Hartree method: Implementation and applications to a Henon-Heiles Hamiltonian
  and to pyrazine}. \emph{Journal of Chemical Physics} \textbf{2011}, \relax
\mciteBstWouldAddEndPunctfalse
\mciteSetBstMidEndSepPunct{\mcitedefaultmidpunct}
{}{\mcitedefaultseppunct}\relax
\EndOfBibitem
\bibitem[Yabana and Bertsch(1996)Yabana, and Bertsch]{Yabana1996}
Yabana,~K.; Bertsch,~G. {Time-dependent local-density approximation in real
  time}. \emph{Physical Review B - Condensed Matter and Materials Physics}
  \textbf{1996}, \relax
\mciteBstWouldAddEndPunctfalse
\mciteSetBstMidEndSepPunct{\mcitedefaultmidpunct}
{}{\mcitedefaultseppunct}\relax
\EndOfBibitem
\bibitem[{De Giovannini} \latin{et~al.}(2013){De Giovannini}, Brunetto, Castro,
  Walkenhorst, and Rubio]{DeGiovannini2013}
{De Giovannini},~U.; Brunetto,~G.; Castro,~A.; Walkenhorst,~J.; Rubio,~A.
  {Simulating pump-probe photoelectron and absorption spectroscopy on the
  attosecond timescale with time-dependent density functional theory}.
  \emph{ChemPhysChem} \textbf{2013}, \relax
\mciteBstWouldAddEndPunctfalse
\mciteSetBstMidEndSepPunct{\mcitedefaultmidpunct}
{}{\mcitedefaultseppunct}\relax
\EndOfBibitem
\bibitem[McLachlan(1964)]{McLachlan1964}
McLachlan,~A.~D. {A variational solution of the time-dependent Schrodinger
  equation}. \emph{Molecular Physics} \textbf{1964}, \relax
\mciteBstWouldAddEndPunctfalse
\mciteSetBstMidEndSepPunct{\mcitedefaultmidpunct}
{}{\mcitedefaultseppunct}\relax
\EndOfBibitem
\bibitem[Vacher \latin{et~al.}(2016)Vacher, Bearpark, and Robb]{Vacher2016}
Vacher,~M.; Bearpark,~M.~J.; Robb,~M.~A. {Direct methods for non-adiabatic
  dynamics: connecting the single-set variational multi-configuration Gaussian
  (vMCG) and Ehrenfest perspectives}. \emph{Theoretical Chemistry Accounts}
  \textbf{2016}, \relax
\mciteBstWouldAddEndPunctfalse
\mciteSetBstMidEndSepPunct{\mcitedefaultmidpunct}
{}{\mcitedefaultseppunct}\relax
\EndOfBibitem
\bibitem[Li \latin{et~al.}(2005)Li, Tully, Schlegel, and Frisch]{Li2005}
Li,~X.; Tully,~J.~C.; Schlegel,~H.~B.; Frisch,~M.~J. {Ab initio Ehrenfest
  dynamics}. \emph{Journal of Chemical Physics} \textbf{2005}, \relax
\mciteBstWouldAddEndPunctfalse
\mciteSetBstMidEndSepPunct{\mcitedefaultmidpunct}
{}{\mcitedefaultseppunct}\relax
\EndOfBibitem
\bibitem[{Andrea Rozzi} \latin{et~al.}(2013){Andrea Rozzi}, {Maria Falke},
  Spallanzani, Rubio, Molinari, Brida, Maiuri, Cerullo, Schramm, Christoffers,
  and Lienau]{AndreaRozzi2013}
{Andrea Rozzi},~C.; {Maria Falke},~S.; Spallanzani,~N.; Rubio,~A.;
  Molinari,~E.; Brida,~D.; Maiuri,~M.; Cerullo,~G.; Schramm,~H.;
  Christoffers,~J. \latin{et~al.}  {Quantum coherence controls the charge
  separation in a prototypical artificial light-harvesting system}.
  \emph{Nature Communications} \textbf{2013}, \relax
\mciteBstWouldAddEndPunctfalse
\mciteSetBstMidEndSepPunct{\mcitedefaultmidpunct}
{}{\mcitedefaultseppunct}\relax
\EndOfBibitem
\bibitem[Krumland \latin{et~al.}(2020)Krumland, Valencia, Pittalis, Rozzi, and
  Cocchi]{Krumland2020}
Krumland,~J.; Valencia,~A.~M.; Pittalis,~S.; Rozzi,~C.~A.; Cocchi,~C.
  {Understanding real-time time-dependent density-functional theory simulations
  of ultrafast laser-induced dynamics in organic molecules}. \emph{The Journal
  of Chemical Physics} \textbf{2020}, \emph{153}, 54106\relax
\mciteBstWouldAddEndPuncttrue
\mciteSetBstMidEndSepPunct{\mcitedefaultmidpunct}
{\mcitedefaultendpunct}{\mcitedefaultseppunct}\relax
\EndOfBibitem
\bibitem[Goings \latin{et~al.}(2016)Goings, Lingerfelt, and Li]{Goings2016}
Goings,~J.~J.; Lingerfelt,~D.~B.; Li,~X. {Can Quantized Vibrational Effects Be
  Obtained from Ehrenfest Mixed Quantum-Classical Dynamics?} \emph{Journal of
  Physical Chemistry Letters} \textbf{2016}, \relax
\mciteBstWouldAddEndPunctfalse
\mciteSetBstMidEndSepPunct{\mcitedefaultmidpunct}
{}{\mcitedefaultseppunct}\relax
\EndOfBibitem
\bibitem[Kapral and Ciccotti(1999)Kapral, and Ciccotti]{Kapral1999}
Kapral,~R.; Ciccotti,~G. {Mixed quantum-classical dynamics}. \emph{Journal of
  Chemical Physics} \textbf{1999}, \relax
\mciteBstWouldAddEndPunctfalse
\mciteSetBstMidEndSepPunct{\mcitedefaultmidpunct}
{}{\mcitedefaultseppunct}\relax
\EndOfBibitem
\bibitem[Broeckhove \latin{et~al.}(1988)Broeckhove, Lathouwers, Kesteloot, and
  {Van Leuven}]{Broeckhove1988}
Broeckhove,~J.; Lathouwers,~L.; Kesteloot,~E.; {Van Leuven},~P. {On the
  equivalence of time-dependent variational principles}. \emph{Chemical Physics
  Letters} \textbf{1988}, \relax
\mciteBstWouldAddEndPunctfalse
\mciteSetBstMidEndSepPunct{\mcitedefaultmidpunct}
{}{\mcitedefaultseppunct}\relax
\EndOfBibitem
\bibitem[Lubich(2004)]{Lubich2004}
Lubich,~C. {On variational approximations in quantum molecular dynamics}.
  \emph{Mathematics of Computation} \textbf{2004}, \relax
\mciteBstWouldAddEndPunctfalse
\mciteSetBstMidEndSepPunct{\mcitedefaultmidpunct}
{}{\mcitedefaultseppunct}\relax
\EndOfBibitem
\bibitem[Ohta(2004)]{Ohta2004}
Ohta,~K. {Time-dependent variational principle with constraints for
  parametrized wave functions}. \emph{Physical Review A - Atomic, Molecular,
  and Optical Physics} \textbf{2004}, \relax
\mciteBstWouldAddEndPunctfalse
\mciteSetBstMidEndSepPunct{\mcitedefaultmidpunct}
{}{\mcitedefaultseppunct}\relax
\EndOfBibitem
\bibitem[Ben(2003)]{Ben-Israel2003}
\emph{Generalized Inverses: Theory and Applications}; Springer New York: New
  York, NY, 2003; pp 201--256\relax
\mciteBstWouldAddEndPuncttrue
\mciteSetBstMidEndSepPunct{\mcitedefaultmidpunct}
{\mcitedefaultendpunct}{\mcitedefaultseppunct}\relax
\EndOfBibitem
\bibitem[Kosloff and Tal-Ezer(1986)Kosloff, and Tal-Ezer]{Kosloff1986}
Kosloff,~R.; Tal-Ezer,~H. {A direct relaxation method for calculating
  eigenfunctions and eigenvalues of the schr{\"{o}}dinger equation on a grid}.
  \emph{Chemical Physics Letters} \textbf{1986}, \relax
\mciteBstWouldAddEndPunctfalse
\mciteSetBstMidEndSepPunct{\mcitedefaultmidpunct}
{}{\mcitedefaultseppunct}\relax
\EndOfBibitem
\bibitem[Shi \latin{et~al.}(2018)Shi, Demler, and {Ignacio Cirac}]{Shi2018}
Shi,~T.; Demler,~E.; {Ignacio Cirac},~J. {Variational study of fermionic and
  bosonic systems with non-Gaussian states: Theory and applications}.
  \emph{Annals of Physics} \textbf{2018}, \relax
\mciteBstWouldAddEndPunctfalse
\mciteSetBstMidEndSepPunct{\mcitedefaultmidpunct}
{}{\mcitedefaultseppunct}\relax
\EndOfBibitem
\bibitem[Kreibich \latin{et~al.}(2001)Kreibich, Lein, Engel, and
  Gross]{Kreibich2001}
Kreibich,~T.; Lein,~M.; Engel,~V.; Gross,~E.~K. {Even-harmonic generation due
  to beyond-born-oppenheimer dynamics}. \emph{Physical Review Letters}
  \textbf{2001}, \relax
\mciteBstWouldAddEndPunctfalse
\mciteSetBstMidEndSepPunct{\mcitedefaultmidpunct}
{}{\mcitedefaultseppunct}\relax
\EndOfBibitem
\bibitem[Lein \latin{et~al.}(2002)Lein, Kreibich, Gross, and Engel]{Lein2002}
Lein,~M.; Kreibich,~T.; Gross,~E.~K.; Engel,~V. {Strong-field ionization
  dynamics of a model H2 molecule}. \emph{Physical Review A - Atomic,
  Molecular, and Optical Physics} \textbf{2002}, \relax
\mciteBstWouldAddEndPunctfalse
\mciteSetBstMidEndSepPunct{\mcitedefaultmidpunct}
{}{\mcitedefaultseppunct}\relax
\EndOfBibitem
\bibitem[Bandrauk and Shon(2002)Bandrauk, and Shon]{Bandrauk2002}
Bandrauk,~A.~D.; Shon,~N.~H. {Attosecond control of ionization and high-order
  harmonic generation in molecules}. \emph{Physical Review A - Atomic,
  Molecular, and Optical Physics} \textbf{2002}, \relax
\mciteBstWouldAddEndPunctfalse
\mciteSetBstMidEndSepPunct{\mcitedefaultmidpunct}
{}{\mcitedefaultseppunct}\relax
\EndOfBibitem
\bibitem[Tancogne-Dejean \latin{et~al.}(2020)Tancogne-Dejean, Oliveira,
  Andrade, Appel, Borca, {Le Breton}, Buchholz, Castro, Corni, Correa, {De
  Giovannini}, Delgado, Eich, Flick, Gil, Gomez, Helbig, H{\"{u}}bener,
  Jest{\"{a}}dt, Jornet-Somoza, Larsen, Lebedeva, L{\"{u}}ders, Marques,
  Ohlmann, Pipolo, Rampp, Rozzi, Strubbe, Sato, Sch{\"{a}}fer, Theophilou,
  Welden, and Rubio]{Tancogne-Dejean2020}
Tancogne-Dejean,~N.; Oliveira,~M.~J.; Andrade,~X.; Appel,~H.; Borca,~C.~H.; {Le
  Breton},~G.; Buchholz,~F.; Castro,~A.; Corni,~S.; Correa,~A.~A.
  \latin{et~al.}  {Octopus, a computational framework for exploring
  light-driven phenomena and quantum dynamics in extended and finite systems}.
  \emph{Journal of Chemical Physics} \textbf{2020}, \relax
\mciteBstWouldAddEndPunctfalse
\mciteSetBstMidEndSepPunct{\mcitedefaultmidpunct}
{}{\mcitedefaultseppunct}\relax
\EndOfBibitem
\bibitem[Gross and Maitra(2012)Gross, and Maitra]{Gross2012}
Gross,~E. K.~U.; Maitra,~N.~T. \emph{{Introduction to TDDFT}}; 2012\relax
\mciteBstWouldAddEndPuncttrue
\mciteSetBstMidEndSepPunct{\mcitedefaultmidpunct}
{\mcitedefaultendpunct}{\mcitedefaultseppunct}\relax
\EndOfBibitem
\bibitem[Koch and Otto(1972)Koch, and Otto]{Koch1972}
Koch,~E.~E.; Otto,~A. {Optical absorption of benzene vapour for photon energies
  from 6 eV to 35 eV}. \emph{Chemical Physics Letters} \textbf{1972},
  \emph{12}, 476--480\relax
\mciteBstWouldAddEndPuncttrue
\mciteSetBstMidEndSepPunct{\mcitedefaultmidpunct}
{\mcitedefaultendpunct}{\mcitedefaultseppunct}\relax
\EndOfBibitem
\bibitem[Gingell \latin{et~al.}(1998)Gingell, Marston, Mason, Zhao, and
  Siggel]{Gingell1998}
Gingell,~J.~M.; Marston,~G.; Mason,~N.~J.; Zhao,~H.; Siggel,~M. R.~F. {On the
  electronic spectroscopy of benzyl alcohol}. \emph{Chemical Physics}
  \textbf{1998}, \emph{237}, 443--449\relax
\mciteBstWouldAddEndPuncttrue
\mciteSetBstMidEndSepPunct{\mcitedefaultmidpunct}
{\mcitedefaultendpunct}{\mcitedefaultseppunct}\relax
\EndOfBibitem
\bibitem[Borges \latin{et~al.}(2003)Borges, Varandas, Rocha, and
  Bielschowsky]{Borges2003}
Borges,~I.; Varandas,~A.~J.; Rocha,~A.~B.; Bielschowsky,~C.~E. {Forbidden
  transitions in benzene}. Journal of Molecular Structure: THEOCHEM. 2003\relax
\mciteBstWouldAddEndPuncttrue
\mciteSetBstMidEndSepPunct{\mcitedefaultmidpunct}
{\mcitedefaultendpunct}{\mcitedefaultseppunct}\relax
\EndOfBibitem
\bibitem[Ridolfi \latin{et~al.}(2020)Ridolfi, Trevisanutto, and
  Pereira]{Ridolfi2020}
Ridolfi,~E.; Trevisanutto,~P.~E.; Pereira,~V.~M. {Expeditious computation of
  nonlinear optical properties of arbitrary order with native electronic
  interactions in the time domain}. \emph{Phys. Rev. B} \textbf{2020},
  \emph{102}, 245110\relax
\mciteBstWouldAddEndPuncttrue
\mciteSetBstMidEndSepPunct{\mcitedefaultmidpunct}
{\mcitedefaultendpunct}{\mcitedefaultseppunct}\relax
\EndOfBibitem
\bibitem[Verlet(1967)]{Verlet1967}
Verlet,~L. {Computer "experiments" on classical fluids. I. Thermodynamical
  properties of Lennard-Jones molecules}. \emph{Physical Review} \textbf{1967},
  \relax
\mciteBstWouldAddEndPunctfalse
\mciteSetBstMidEndSepPunct{\mcitedefaultmidpunct}
{}{\mcitedefaultseppunct}\relax
\EndOfBibitem
\bibitem[de~Boor(2001)]{DeBoor2001}
de~Boor,~C. \emph{Springer-Verlag, New York}; 2001\relax
\mciteBstWouldAddEndPuncttrue
\mciteSetBstMidEndSepPunct{\mcitedefaultmidpunct}
{\mcitedefaultendpunct}{\mcitedefaultseppunct}\relax
\EndOfBibitem
\bibitem[Fairlie(1999)]{FAIRLIE1999365}
Fairlie,~D.~B. {Moyal brackets, star products and the generalised Wigner
  function}. \emph{Chaos, Solitons {\&} Fractals} \textbf{1999}, \emph{10},
  365--371\relax
\mciteBstWouldAddEndPuncttrue
\mciteSetBstMidEndSepPunct{\mcitedefaultmidpunct}
{\mcitedefaultendpunct}{\mcitedefaultseppunct}\relax
\EndOfBibitem
\bibitem[McKemmish \latin{et~al.}(2011)McKemmish, McKenzie, Hush, and
  Reimers]{McKemmish2011}
McKemmish,~L.~K.; McKenzie,~R.~H.; Hush,~N.~S.; Reimers,~J.~R. {Quantum
  entanglement between electronic and vibrational degrees of freedom in
  molecules}. \emph{Journal of Chemical Physics} \textbf{2011}, \relax
\mciteBstWouldAddEndPunctfalse
\mciteSetBstMidEndSepPunct{\mcitedefaultmidpunct}
{}{\mcitedefaultseppunct}\relax
\EndOfBibitem
\bibitem[Yabana \latin{et~al.}(2006)Yabana, Nakatsukasa, Iwata, and
  Bertsch]{Yabana2006}
Yabana,~K.; Nakatsukasa,~T.; Iwata,~J.-I.; Bertsch,~G.~F. {Real-time,
  real-space implementation of the linear response time-dependent
  density-functional theory}. \emph{physica status solidi (b)} \textbf{2006},
  \emph{243}, 1121--1138\relax
\mciteBstWouldAddEndPuncttrue
\mciteSetBstMidEndSepPunct{\mcitedefaultmidpunct}
{\mcitedefaultendpunct}{\mcitedefaultseppunct}\relax
\EndOfBibitem
\bibitem[Muga \latin{et~al.}(2004)Muga, Palao, Navarro, and
  Egusquiza]{Muga2004}
Muga,~J.~G.; Palao,~J.~P.; Navarro,~B.; Egusquiza,~I.~L. {Complex absorbing
  potentials}. 2004\relax
\mciteBstWouldAddEndPuncttrue
\mciteSetBstMidEndSepPunct{\mcitedefaultmidpunct}
{\mcitedefaultendpunct}{\mcitedefaultseppunct}\relax
\EndOfBibitem
\end{mcitethebibliography}

\end{document}